%% file: main.tex
\documentclass[sigconf]{acmart}
\usepackage{multirow}
\usepackage{enumitem}

\AtBeginDocument{%
  }

\copyrightyear{2026}
\acmYear{2026}
\setcopyright{cc}
\setcctype{by-nc-nd}
\acmConference[CCS '26]{Proceedings of the 2026 ACM SIGSAC Conference on Computer and Communications Security}{November 15--19, 2026}{The Hague, Netherlands}
\acmBooktitle{Proceedings of the 2026 ACM SIGSAC Conference on Computer and Communications Security (CCS '26), November 15--19, 2026, The Hague, Netherlands}
\acmDOI{10.1145/3830454.3846705}
\acmISBN{979-8-4007-2871-6/2026/11}

\begin{document}

\title[It Takes a Forum: It Takes a Forum: Resources \& Stakeholders Sustaining the Online SNCII Ecosystem]{It Takes a Forum: Resources \& Stakeholders Sustaining the Online Synthetic Non-Consensual Intimate Image Creation Ecosystem}





\author{Bernardo B. P. Medeiros}
\authornote{These authors contributed equally to this work.}
\email{b.broetopontimed@ufl.edu}
\affiliation{
  \institution{University of Florida}
  \city{Gainesville}
  \state{FL}
  \country{USA}
}

\author{Malvika Jadhav}
\authornotemark[1]
\email{jadhav.m@ufl.edu}
\affiliation{
  \institution{University of Florida}
  \city{Gainesville}
  \state{FL}
  \country{USA}
}

\author{Allison Lu}
\email{allison.lu@ufl.edu}
\affiliation{
  \institution{University of Florida}
  \city{Gainesville}
  \state{FL}
  \country{USA}
}

\author{Tadayoshi Kohno}
\email{yoshi.kohno@georgetown.edu}
\affiliation{
  \institution{Georgetown University}
  \city{Washington}
  \state{D.C.}
  \country{USA}
}

\author{Vincent Bindschaedler}
\email{vbindschaedler@ufl.edu}
\affiliation{
  \institution{University of Florida}
  \city{Gainesville}
  \state{FL}
  \country{USA}
}

\author{Kevin R. B. Butler}
\email{butler@ufl.edu}
\affiliation{
  \institution{University of Florida}
  \city{Gainesville}
  \state{FL}
  \country{USA}
}

\renewcommand{\shortauthors}{Bernardo B. P. Medeiros et al.}

\begin{abstract}
\input{Sections/abstract}

\vspace{1em}
\noindent\fbox{%
    \parbox{\dimexpr\linewidth-2\fboxsep-2\fboxrule}{%
        \textcolor{red}{\textbf{Content Warning:} This paper discusses synthetic non-consensual intimate image creation communities and contains content that some might consider sensitive.}
    }%
}
\end{abstract}



\maketitle

\input{Sections/intro}

\input{Sections/background}

\input{Sections/methodology}

\input{Sections/results}

\input{Sections/discussion}

\input{Sections/related_work}

\input{Sections/conclusion}

\input{Sections/ethics}

\begin{acks}
We would like to thank Natalie Grace Brigham, Elissa Redmiles, and Jaron Mink for their invaluable input. This work was supported in part by the National Science Foundation (CNS-2055123, CNS-2206950, and CNS-2205171). Tadayoshi Kohno is supported by the Robert L. McDevitt, K.S.G., K.C.H.S. and Catherine H. McDevitt L.C.H.S. Chair in Computer Science at Georgetown University. Any opinions, findings, conclusions, or recommendations expressed in this material are those of the authors and do not necessarily reflect the views of the National Science Foundation.
\end{acks}

\bibliographystyle{ACM-Reference-Format}
\bibliography{ref}

\appendix
\input{Sections/open-science}
\input{Sections/appendix}

\end{document}

%% file: Sections/abstract.tex
\label{abstract}
Many malicious actors responsible for disseminating synthetic non-consensual intimate imagery (SNCII) operate within internet forums to exchange resources, strategies, and generated content across multiple platforms. Technically-sophisticated actors gravitate toward certain communities (e.g., 4chan), while lower-sophistication end-users are more active on others (e.g., Reddit). To characterize key stakeholders in the broader ecosystem, we perform an integrated analysis of multiple communities, analyzing 282,154 4chan comments and 78,308 Reddit submissions spanning 165 days between June and November 2025 to characterize involved actors, actions, and resources. We uncover a litany of resources both shared and discussed by myriad actors on both forums, including 156 low-ranking adaptation (LoRA) files used to finetune models to target 141 female celebrities (on 4chan), and links to 54 commercial nudification applications (on Reddit). Our results illustrate: (a) that users with differing levels of technical sophistication employ and share a wide range of primary resources facilitating SNCII content creation as well as numerous secondary resources facilitating dissemination; and (b) that knowledge transfer between experts and newcomers facilitates propagation of these illicit resources. Based on our empirical analysis, we identify gaps in existing SNCII regulatory infrastructure and synthesize several critical intervention points for bolstering deterrence.

%% file: Sections/intro.tex
\section{Introduction}
Online actors are producing synthetic non-consensual intimate imagery (SNCII) of real persons at scale. Since 2023, nearly 90 schools across 28 countries have reported incidents where students used AI ``nudify'' tools to produce sexual imagery of classmates. The National Center for Missing \& Exploited Children (NCMEC) received over 30,000 CyberTipline reports in 2025 of users attempting to generate child sexual abuse material (CSAM) by uploading images by using text prompts~\cite{missingkidsWorkNever}.  At least 600 victims have been identified, with a single North American case involving more than 60 alleged victims~\cite{techbuzzDeepfakeNudes}. Recent laws aimed at SNCII, such as the U.S. TAKE IT DOWN Act, the EU AI Act, and the UK Online Safety Act~\cite{sen_cruz_s146_2025, hart_eu_2026, OnlineSafetyAct}, have largely been shaped by high-profile cases involving public figures and incidents like the recent misuse of xAI’s Grok~\cite{cress_ofcom_2026,montgomery_taylor_2024}. However, these existing SNCII laws overlook a critical part of the problem: the underlying tools and systems that make this content so easy to create in the first place. Tools and systems span commercial ``nudify'' applications promoted on internet platforms, open-source model variants and fine-tuning adapters circulated on community forums, file-sharing services used to distribute training data and model weights, and the people who teach newcomers how to assemble these components into working pipelines. Even more comprehensive laws criminalizing image generation, such as the UK's Online Safety Act~\cite{OnlineSafetyAct}, leave resource-distributing stakeholders undeterred.

Past research has investigated SNCII creation tools~\cite{hawkins2025deepfakes, gibson2025analyzing, ding_malicious_2025}, both commercial~\cite{gibson2025analyzing} and open-source~\cite{hawkins2025deepfakes}, and some research has characterized communities on a ``deepfake website''~\cite{han2025characterizing}. However, to date, no work has deeply investigated SNCII creation and dissemination from a resource-centric perspective: the resources people share, the \emph{knowledge providers} (actors such as educators who pass on procedural knowledge through tutorials, curated guides, and interactive troubleshooting), the \emph{tool providers} (actors who build or distribute creation tools, from commercial nudification app operators to users who monetize models and adapters), and the channels through which sharing occurs.

This work characterizes resources and stakeholders by analyzing data from 4chan and Reddit forums where creators, educators, and other ecosystem stakeholders congregate.\footnote{Our focus in this paper is strictly on AI-fabricated {\em non-consensual} sexual imagery of {\em real persons}, therefore we exclude (for example) AI-generated fakes of fictional characters, which are prevalent on Reddit.} These platforms play distinct roles in the ecosystem: 4chan hosts technical resource sharing and knowledge transfer among creators, while Reddit shows a prevalence of promotional posts for commercial nudification tools. Our analysis employs a dual-pipeline methodology to examine each platform's community for the activity it uniquely surfaces. Our goal is to understand the resource-sharing framework underlying these communities, which is crucial for informing SNCII mitigation. 

Therefore, we focus on the following research questions:

\begin{enumerate}[label=(\arabic*), leftmargin=2.5em]
\item What resources are shared in SNCII creation communities? \label{rq:rq1}
\item What role do stakeholders play in the resource sharing ecosystem, and which stakeholders are most influential?\label{rq:rq2}
\item How do knowledge providers facilitate SNCII creation? \label{rq:rq3}
\item How do tool providers enable access to SNCII creation? \label{rq:rq4}
\end{enumerate}

To answer these questions, we analyze 282,154 4chan posts collected during the period from June 9th 2025 to November 21st 2025, and 78,308 Reddit submissions (i.e., posts and comments) collected from six subreddits during the same time period.\footnote{We end data collection when the 4chan community we study transitions to another board that is archived on another, considerably more difficult to scrape 4chan archiving website. See Appendix~\ref{sec:open} for more details.} After developing a multi-label behavioral codebook of actors, actions, and resources through in-depth manual analyses of a random sample of 200 posts from each dataset until saturation, we leverage our codebook to characterize resource-sharing behaviors at scale. We elaborate on this methodology, including limitations, in Section~\ref{sec:methodology}. 

Our study's stakeholder-centric analysis of SNCII ecosystems reveals complex interactions underlying resource-sharing practices within communities catering to users with differing technical sophistication levels. Key findings include:

\begin{itemize}[nosep]
\begin{sloppypar}
\item \textbf{A characterization of ecosystem resources and stakeholders}, including: (1) a codebook identifying stakeholders (i.e., actor types) involved in community and the resources those stakeholders rely on, including both `primary resources' used for nudification (LoRAs, models, prompts, UIs) and `secondary resources' used for distribution (file-sharing services, pastebin guides, messaging platforms); and (2) an analysis at scale, leveraging our 4chan and Reddit datasets to quantitatively investigate the role of stakeholders in resource dissemination and identify which resources stakeholders rely on. Our work is the first to characterize this ecosystem on general-purpose internet forums.
\end{sloppypar}

\item \textbf{A shadow infrastructure} that adapts to enforcement, which facilitates the sharing of model customization resources (e.g., tutorials, base models, LoRAs) through third-party ephemeral file-sharing services or through private messaging applications such as Telegram and Session. Primary resources move through secondary resource channels, which redirects users to less restricted services. Furthermore, platforms have complementary roles: 4chan supports technical resource sharing, while Reddit hosts commercial tool information and referrals.

\item \textbf{\textsc{Educators}} acting as bridges that help newcomers transition into active SNCII creators, whose impact extends beyond tutorials to include interactive troubleshooting, resource curation, and follow-up guidance, are linked to 56.3\% of creator transitions. Our analysis shows that these actors are identifiable at scale.

\item \textbf{An accountability gap} indicated by the fact that  many community members display familiarity with regulations, but are unconcerned about their effects on their SNCII creation and sharing activities. Despite this, real-person likeness policy of Civitai, one of the major public repositories of user-trained AI image-generation models~\cite{hawkins2025deepfakes}, shows that effective action by hosting services is feasible, and the impact can be measured.
\end{itemize}

\noindent Synthesizing these findings, we identify possible intervention points, including mitigating educator influence, leveraging infrastructure chokepoints, and promoting cross-platform moderation.

\paragraph{Structure of the Paper.}
Section~\ref{sec:background} provides background on SNCII creation and motivation for our study. The data collection pipeline is described in Section~~\ref{sec:collection}. Section~\ref{sec:methodology} provides a study methodology overview, with further details linked in Appendix~\ref{sec:open}. Section~\ref{sec:results} presents the findings of our analyses of both forum communities, along with several case studies. We discuss the broader context surrounding these findings and related takeaways in Section~\ref{sec:discussion}. Section~\ref{sec:rel-work} provides a summary of related work. Section~\ref{sec:concl} concludes.

%% file: Sections/background.tex
\section{Background \& Motivation} 
\label{sec:background}
Synthetic non-consensual intimate images (SNCII) are computer-created images of real individuals in intimate contexts generated without their consent. A variety of tools can be used for the creation of this content, from traditional graphics editors (e.g., Adobe Photoshop) to cutting-edge generative AI models (both web-hosted commercial models and locally-hosted open-source models).

Recent reports document how distinct stakeholder groups contribute to the production and dissemination of SNCII. Commercial nudification tools promoted on mainstream platforms have reached millions of users, with investigations identifying over 100 such applications on Apple and Google app stores~\cite{murti_apple_2026}. At the same time, technical forums exhibit substantial activity spikes following media coverage of tools such as DeepNude~\cite{puglielli_ecosystem_2025}. Reported incidents illustrate that both commercial distribution and community-level knowledge sharing can translate into tangible offline harms~\cite{tenbarge_beverly_2024}. These cases suggest that SNCII harms are not confined to a single platform or actor type, but instead arise from an interconnected ecosystem. A consistent enabling factor across contexts is the online circulation of resources, including commercial tools, open source models, and procedural guidance for generating SNCII~\cite{ding_malicious_2025}. 

Some past work has characterized nudification resources and the broader ecosystem~\cite{kamachee_video_2026, ding_malicious_2025, gibson2025analyzing, hawkins2025deepfakes, han2025characterizing, cuevas2026deepfakepornographyresilientregulatory}. Resources include not only ``nudification applications,'' which are low-barrier-to-entry SNCII creation tools typically hosted on webapps or through Telegram bots that allow users to ``nudify'' input images by using paid credits~\cite{bharadia_millions_2026, gibson2025analyzing}, but also open-source model variants trained to optimize nudification outputs. Open-source model variants are supplemented by low-ranking adapters (LoRAs), which can be applied over top of pre-trained models to adapt them to particular contexts. LoRAs can be effective with as few as 20 training images and have very low VRAM requirements, making them uniquely accessible to users with limited compute power~\cite{hawkins2025deepfakes}. 

Research characterizing the ecosystem identifies three broad areas for content distribution: ``deepfake websites'' (e.g., the now-defunct MrDeepfakes)~\cite{han2025characterizing}, social media platforms (e.g., X, YouTube, TikTok), and online forums (e.g., Reddit, 4chan)~\cite{ding_malicious_2025}. These studies establish where SNCII circulates, but not how the capability to produce it is supplied. Han et al. ~\cite{han2025characterizing} characterize a dedicated marketplace. Ding and Suresh ~\cite{ding_malicious_2025} map the technical components enabling SNCII creation and the governance mechanisms that fail to reach them, at the level of the technology stack rather than the actors who move those components between users. Neither measures what circulates on general-purpose forums, which actors supply it, or how newcomers acquire the capability to create SNCII themselves. Our work fills this gap through RQ(1)--RQ(4), presenting the first \textit{stakeholder-centric} characterization of nudification activity on \textit{general-purpose forums}. By characterizing critical points in the ecosystem, our analysis seeks to inform SNCII mitigation efforts. 

%% file: Sections/methodology.tex
\input{tables/stakeholders.tex}

\section{Data Collection}\label{sec:collection}
To comprehensively characterize ecosystem behaviors, we gathered and analyzed submissions (i.e., posts and comments) from two internet forums: 4chan and Reddit. These platforms serve different functions within the SNCII ecosystem~\cite{ding_malicious_2025} and produce structurally different data: 4chan consists of short, contextually dense posts organized around direct reply chains within a single dedicated thread, while Reddit submissions are longer, self-contained, and heterogeneous in topic. See Appendix~\ref{sec:open} for more details.

\input{tables/resources_direct}

\textit{4chan.}
We searched a 4chan archive website for threads on the ``/b/'' board, which hosts a persistent community thread dedicated to SNCII content creation and resource sharing. Using Beautiful Soup ~\cite{richardson2007beautiful} we scraped 813 unique thread URLs spanning June 9th\textendash November 21st 2025 (165 days, avg \textasciitilde5 threads/day) for a total of 282,154 comments, including 12,069 by 391 non-anonymous users. 4chan users are anonymous by default and usernames are opt-in; the platform has no private messaging, and all posts are public. This anonymity-by-default property shapes nearly all of our 4chan analyses, since the overwhelming majority of activity cannot be attributed to persistent identities. Of the  282,154 comments, 119,418 contained images or videos, identified through HTML metadata. Images were not viewed or stored as part of our commitment to ethical research practices, given the harmful nature of this content (\nameref{sec:ethics}). Refer to Appendix~\ref{sec:open} for additional data collection information.

\textit{Reddit.} We gathered 27,689 posts and 50,619 comments by 17,386 unique authors from six NSFW AI subreddits over the same period (June 9th to November 21st, 2025), to ensure temporal alignment, using Arctic Shift~\cite{heitmann2026arcticshift}. We selected the subreddits based on their stated focus on NSFW AI content. 

In total, our analysis covers 282,154 4chan comments and 78,308 Reddit submissions across 165 days.

\section{Methodology}\label{sec:methodology}

\label{subsec:design}
Building on prior work characterizing the SNCII ecosystem~\cite{ding_malicious_2025}, we present the first empirical characterization of resource sharing practices on internet forums. See \nameref{sec:ethics} for a discussion of the steps taken to ensure ethical investigation throughout the research process.

\begin{figure}[t]
    \centering
    \includegraphics[width=\columnwidth]{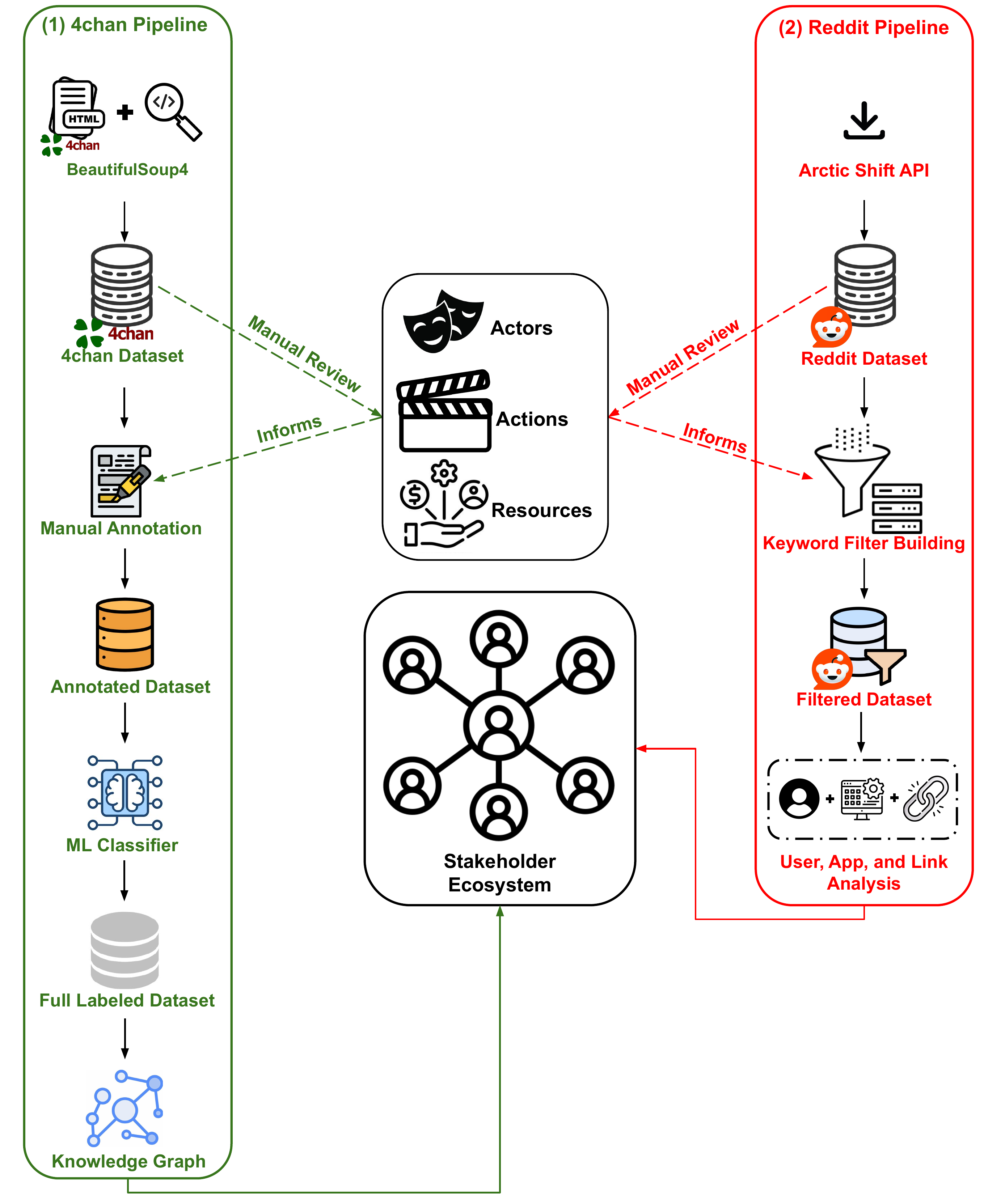}
    \caption{Overview of our dual-pipeline methodology. \textit{(1) 4chan pipeline:} data is collected via web scraping, filtered, manually annotated, then used to train a transformer-based classifier to label the full dataset for knowledge graph construction. \textit{(2) Reddit pipeline:} data is downloaded via Arctic Shift, filtered for commercial nudification application mentions, checked for referral link patterns, and analyzed at the user and app levels. Both datasets inform a shared multi-label behavioral codebook of actors, actions, and resources. Results jointly characterize the broader ecosystem.}
    \label{fig:pipeline}
\end{figure}

\subsection{Codebook Development}
\label{subsec:typology}
Through iterative analysis of 200 samples from each platform informed by prior analyses of online communities~\cite{brandtzaeg2010towards,bernstein20114chan,timmerman2023studying}, we developed a multi-label behavioral codebook of actors, actions, and resources (Tables~\ref{tab:typology} and~\ref{table:resources}). Codes were assigned at the post level, recording what a post does rather than a fixed attribute of its author. A single post may carry multiple actor, action, and resource labels. A post sharing a LoRA link and praising another user's output, for example, is coded both \textsc{share\_resource} and \textsc{react\_praise} and counted under each, which is why label counts exceed the number of posts (Table ~\ref{table:silver_labels}). The codebook's structure is hierarchical: each post has actor, action, resource, and harm-magnifier codes which are not mutually exclusive and collectively represent post-level content (see Appendix~\ref{sec:open}). Actor codes record the target of the activity, distinguishing public figures, private individuals, and unspecified targets, and carry a veteran modifier for posters with a username or demonstrated historical community knowledge. Action codes record the action being taken by an actor (e.g., socializing, sharing a resource, reacting to a shared image). Resource codes are annotated on two axes: the resource type (e.g., LoRA, model, file-sharing service) and the manner in which it is shared, whether named in passing, linked directly, accompanied by instructions, or reviewed. A harm-magnifier axis flags posts that reference minors or target private individuals. Each annotated post therefore carries a set of action codes, a set of actor codes, any resources it references, and any harm magnifiers, alongside its metadata and whether it contains an image; Table ~\ref{tab:kg-schema} lists these as the attributes stored per post. 

Preliminary exploration of our collected data revealed platform differences: 4chan hosted more technical discussion and content sharing, while Reddit activity skewed toward commercial tool promotion. For instance: the most-mentioned nudification checkpoint on 4chan (1,049 mentions) saw only 18 mentions on Reddit (1.7\% of 4chan mentions). These differences motivated our dual-pipeline methodology (Figure~\ref{fig:pipeline}): 4chan threads were densely structured around SNCII-specific content, while Reddit submissions were more heterogeneous. See Appendix~\ref{sec:open} for more detail.

Key actors identified in our analysis included \textbf{\textsc{genners}}, who \textsc{share} generated SNCII content and \textsc{respond} to requests for such content. \textbf{\textsc{Educators}} primarily \textsc{respond} to \textsc{requests} for resources and \textsc{share} resources. \textbf{\textsc{Socializers}} \textsc{discuss}, \textsc{repost}, and \textsc{react} to both generated images and shared resources, often engaging in conversations related to SNCII content creation but not specific resources. \textbf{\textsc{Requesters}} are chiefly responsible for \textsc{requesting} resources or generated SNCII content. \textbf{\textsc{Promoters}} contribute by \textsc{advertising} nudification products. We define \textbf{\textsc{veterans}} as long-time community members who engage in \textsc{meta} discussion and/or have a username (on 4chan). We applied the \textsc{Veterans} label only on 4chan, since the Reddit subreddits we examined contain little \textsc{Meta} commentary, and influence analysis showed that individual posters do not tend to be highly influential on our target subreddits, although \textsc{Veterans} are not by definition excluded from appearing on Reddit (see Section~\ref{subsec:promoters}). Actor types were unevenly distributed across platforms: \textsc{Genners} and \textsc{Educators} were most active on 4chan, while \textsc{Promoters} were more prominent on Reddit, where the submissions in our dataset were frequently commercial in nature. 

\paragraph{Resource Characterization}
In characterizing the nudification resources disseminated on online forum communities (``primary resources''), we uncovered resources used for distribution (``secondary resources''). Secondary resources included: file sharing services like GoFile, which are used to share datasets for LoRA training and/or LoRA files themselves; model hosting websites like Civitai and HuggingFace, which are used to share both base models and finetuning layers; personal blogs and other similar websites (e.g., Rentry\footnote{Rentry is a pastebin service that hosts formatted text rather than binary files. In this ecosystem, Rentry guides typically consist of curated collections of links and step-by-step instructions for setting up nudification pipelines, whereas file-sharing services such as GoFile and/or torrenting services are used for hosting files (e.g., model checkpoints, LoRAs, datasets).}) which are frequently used to host instructions or guides; and illicit groups on messaging platforms like Telegram and Session, which can be used for resource and content sharing. We describe relevant resources in more detail in Table~\ref{table:resources}.

\begin{table}[t]
    \centering
    \small
    \begin{tabular}{@{}lr@{}}
        \toprule
        \textbf{Post Node Attributes} & \\
        \midrule
        Base & \texttt{post\_id}, \texttt{username}, \texttt{timestamp}, \\
             & \texttt{text}, \texttt{has\_image}\\
        Labels & \texttt{actions[]}, \texttt{actors[]}, \\
               & \texttt{resources[]}, \texttt{harm\_signals[]} \\
        \midrule
        \textbf{Edges (Post $\rightarrow$ Post)} & \textbf{Count} \\
        \midrule
        \textsc{discusses} & 216,520 \\
        \textsc{reacts\_to} & 135,187 \\
        \textsc{responds\_to} & 3,351 \\
        \bottomrule
    \end{tabular}
    \caption{\textit{[4chan]} Knowledge graph schema and statistics.}
    \label{tab:kg-schema}
\end{table}

\begin{table}[t!]
\centering
\small
\begin{tabular}{@{}lcc@{}}
\toprule
\textbf{Category} & \textbf{Macro F1} & \textbf{95\% CI} \\
\midrule
Resources     & 0.92 & [0.89, 0.96] \\
Harm Magnifiers  & 0.92 & [0.88, 0.95] \\
Actions       & 0.68 & [0.62, 0.74] \\
Actors        & 0.85 & [0.81, 0.87] \\
\bottomrule
\end{tabular}
\caption{\textit{[4chan]} Silver dataset validation against 200 samples not in the Golden dataset. We report macro-averaged F1 scores with 95\% bootstrap confidence intervals.}
\label{tab:silver_validation}
\end{table}

\subsection{4chan}

\textit{Manual Analysis.} \label{subsec:manual} We selected the 400 posts for annotation using a stratified sampling strategy that prioritized posts mentioning resources identified during preliminary manual exploration, so that the annotated set is dense in resource discussion. Two researchers independently annotated each post, adding new codes whenever a post could not be captured by the existing set, and stopped at saturation: no new codes emerged, and every post was coded under the final codebook. Our released codebook (Appendix ~\ref{sec:open}) gives per-code inclusion and exclusion criteria with examples. Divergent codes were resolved through discussion, with a third annotator settling any disagreements that remained to produce the gold dataset. Inter-rater reliability (IRR) before resolution was 0.71 ~\cite{kupper1989assessing}, comparable to prior work in this space (0.72 ~\cite{han2025characterizing}). The IRR score understates agreement because Kupper--Hafner, which is used to calculate agreement for multi-label codebooks, treats partial agreement on multi-label posts as disagreement~\cite{kupper1989assessing}.

\textit{Automated Labeling.}
\label{subsec:automated}To scale analysis beyond our 400 manual annotations to the full 4chan corpus with 282,154 posts, we developed a hybrid pipeline. Actor and action labels require interpreting communicative intent (e.g., ``LoRA training'' could indicate an \textsc{educator} sharing instructions or a \textsc{requester} seeking help), so we trained SetFit~\cite{tunstall2022efficient}, a few-shot sentence-transformer classifier. Resources and harm signals have recognizable surface cues (tool names, URLs, jargon), so we used rule-based keyword matching (see Appendix~\ref{sec:open} for more detail). We validated the pipeline on 200 held-out samples (Table~\ref{tab:silver_validation}): SetFit achieved F1=0.85 for actors and F1=0.68 for actions; keyword matching achieved F1=0.92 for both resources and harm magnifiers. See Section~\ref{subsec:scope-limitations} for further discussion.

\begin{figure}[t]
    \centering
    \includegraphics[width=0.48\textwidth]{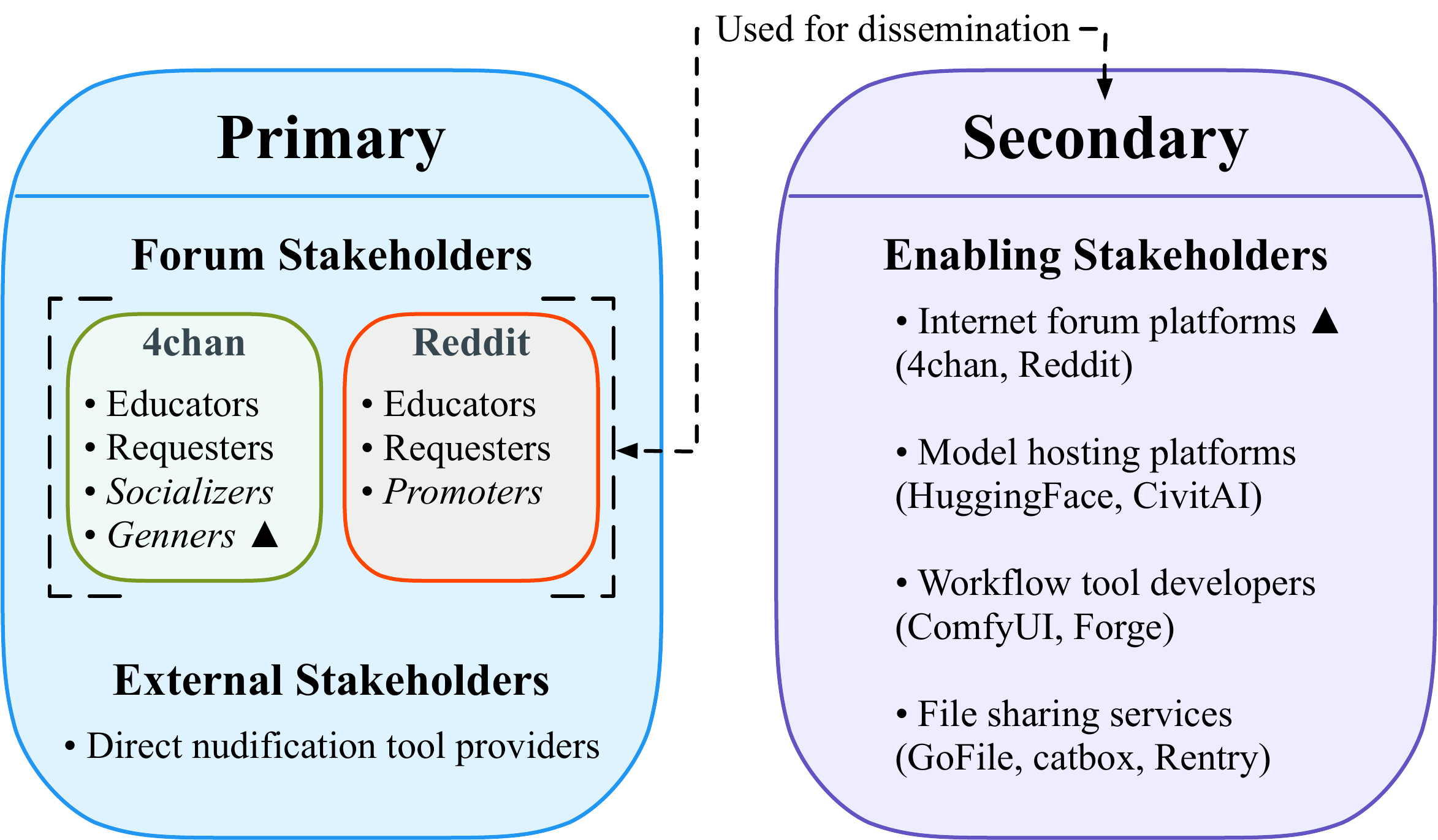}
    \caption{Primary and secondary ecosystem stakeholders. \textit{Primary stakeholders} directly participate in the nudification ecosystem. \textit{Secondary stakeholders} enable primary resource distribution. Stakeholders only identified on one forum are shown in \textit{italics}. \ensuremath{\blacktriangle} designates stakeholders held accountable by existing U.S. legal infrastructure. Most stakeholders are not held accountable: only those who post SNCII content online and the platforms hosting that content are covered.}
    \label{fig:stakeholder}
\end{figure}

\textit{Knowledge Graph.}
\label{subsec:KG}
We constructed a Neo4j knowledge graph from the labeled corpus, with posts as nodes carrying metadata and labels as attributes. Edges encode reply-based interactions using 4chan's markup (\texttt{>>post\_id}). This structure enables analysis of interaction patterns and influence propagation across community response chains. Table~\ref{tab:kg-schema} summarizes the schema, with additional detail available at Appendix~\ref{sec:open}.

\input{tables/silver_labels_table}

\subsection{Reddit}
Reddit’s data properties made it unsuitable for the same actor and action classification pipeline used for 4chan. Identifying actors such as \textsc{educators}, \textsc{genners}, \textsc{socializers}, and others requires interpreting communicative intent: whether a post about LoRA training reflects someone sharing instructions or someone seeking help, for instance, is only legible when that post is read in the context of a short exchange embedded in a topically-focused reply chain. The Reddit submissions we analyzed, on the contrary, were long, self-contained, and covered a wide range of topics within each subreddit. They also included large amounts of AI-generated sexual content involving fictional characters, falling outside our threat model. As a result, the contextual cues that make actor identification feasible on 4chan are dispersed across multiple sentences in a long Reddit post. A classifier trained on 4chan discourse did not transfer well to this environment, and achieving reliable annotations on Reddit would require a much larger, fundamentally different effort due to its noise and heterogeneity. 

The behavior that remained consistently identifiable on Reddit is promotion of commercial nudification apps, where app names and referral links provided clear signals that generalize across subreddits. Because these commercial actors (\textsc{Promoters}) play a central role in the SNCII ecosystem, both in enabling access to and sustaining the distribution of these resources, capturing their activity is essential to understanding the broader system. Our analysis was therefore scoped accordingly, given Reddit's structural properties.

\textit{Manual Analysis.} Building on standard practice in Reddit content analysis research~\cite{oak2025victims, floyd2021understanding, wu2022conversations}, we used a keyword-based filtering strategy. The research team reviewed 400 Reddit submissions containing hyperlinks until saturation, identifying the 54 commercial nudification apps that comprised our keyword filter. See Appendix~\ref{sec:open} for additional detail and Section~\ref{subsec:scope-limitations} for discussion of preprocessing limitations. 

\textit{Automated Analysis.}
We applied the keyword filter to identify posts mentioning commercial nudification apps, and searched for referral code patterns in hyperlinks. After filtering, we were left with 3,683 Reddit submissions mentioning a commercial nudification application (4.7\% of the initial dataset). To validate our filtering strategy, two researchers each manually reviewed 200 randomly sampled posts from the filtered dataset. Nearly all samples (99.8\%, 399 out of 400) mentioned a commercial nudification app, with only one false positive, indicating that filtering effectively removed irrelevant posts. Beyond subreddit-level trends, we conducted application-level analysis to track app mentions over time, as well as user-level analysis by mapping referral links to individual users and their posting behavior. More detail is provided in Appendix~\ref{sec:open}.

\subsection{Methodological Limitations}
\label{subsec:scope-limitations}
Four properties of our data shape the interpretation of our results. We analyze text and never images, so our evidence covers what participants share and say about resources, not the content those resources generate. Anonymity is the default on 4chan, and with over 95\% of posts carrying no persistent identity, we characterized influential activity in aggregate alongside the small subset of named accounts rather than tracing individual histories (Section~\ref{subsec:influence}). Additionally, 4chan comments are highly contextual; therefore, our automated annotation model may underestimate implicitly mentioned entities (e.g., LoRAs). Finally, our collection window is fixed, and users who were already creating content before it opened (June 9th) appear as creators in their first observed post, such that the role transition rate we report is a lower bound (Section~\ref{subsec:knowledge-transfer}). 

For 4chan, the F1 score of 0.68 obtained when training the ``action'' classifier reflects the difficulty of applying labels reliably at scale and falls within the range reported for comparable security-NLP tasks. A recent SoK reports a best F1 of 0.71 for this problem~\cite{buchel2025sok}, and domain-specific classifiers typically demonstrate F1 scores of 0.6--0.7 for similar tasks~\cite{orbinato2022automatic}. Furthermore, most classifier errors are missed labels rather than false positives, ensuring the counts we report are conservative lower bounds. For Reddit, because our filtering strategy prioritizes precision over recall, it is possible we have not captured all app mentions. 

These properties of our datasets and analysis pipeline ensure that the quantitative results reported in Section~\ref{sec:results} are conservative lower bounds of the true scale of activity in the observed ecosystem. We discuss broader limitations in Section~\ref{sec:discussion}.

%% file: tables/stakeholders.tex
\begin{table}[t!]
\centering
\small
\begin{tabular}
{@{}p{0.17\linewidth}p{0.77\linewidth}@{}}
\toprule
\textbf{ACTOR} & \textbf{ROLE} \\
\midrule
\textsc{Genner} & Share generated content, respond to content requests. \\
\textsc{Educator} & Share resources, respond to resource requests. \\
\textsc{Socializer} & Discuss, repost, and react to shared content/resources; engage in general conversation. \\
\textsc{Requester} & Request resources and/or generated content. \\
\textsc{Promoter} & Advertise nudification products. \\
\textsc{Veteran} & Long-time member; engage in meta discussion. \\
\midrule
\textbf{ACTION} & \textbf{DEFINITION} \\
\midrule
\textsc{Advertise} & Promote a nudification product or service. \\
\textsc{Discuss} & Engage in general SNCII-related conversation. \\
\textsc{Meta} & Engage in discussion about the community itself. \\
\textsc{React} & Respond affectively to shared content/resources. \\
\textsc{Repost} & Reshare content originally posted by another user. \\
\textsc{Request} & Solicit resources or generated content. \\
\textsc{Respond} & Reply to a request for resources or generated content. \\
\textsc{Share} & Disseminate a resource or generated SNCII content. \\
\bottomrule
\end{tabular}
\caption{Multi-label behavioral codes for actors and actions involved in the nudification resource-sharing ecosystem on internet forums. Labels are not mutually exclusive.}
\label{tab:typology}
\end{table}

%% file: tables/resources_direct.tex
\begin{table}[t]
\centering
\small
\begin{tabular}{@{}p{2.1cm}p{5.5cm}@{}}
\toprule
\textbf{RESOURCE} & \textbf{DESCRIPTION} \\
\midrule
\multicolumn{2}{@{}l@{}}{\textsc{Primary Resources}} \\
\midrule
Model & Diffusion model checkpoints for SNCII creation. \\
LoRA & Fine-tuning layers for training models on targets. \\
Dataset & Target datasets shared for creating LoRAs. \\
UI & Interfaces (e.g., ComfyUI, Forge) for model use. \\
Prompt & Text inputs for image generation. \\
Hardware Instr. & Model memory requirements. \\
Expert Genner & Prolific content sharers. \\
Nudif. Product & Commercial tools/bots. \\
Credit Incentive & Tokens/credits for commercial apps. \\
\midrule
\multicolumn{2}{@{}l@{}}{\textsc{Secondary Resources}} \\
\midrule
File Sharing & Platforms for hosting files (e.g., catbox.moe). \\
Model Hosting & Platforms for sharing models (e.g., HuggingFace). \\
Personal Blog & Individual user websites. \\
Rentry Guide & Instruction guides on Rentry. \\
Telegram & Messaging app for sharing. \\
Web Archive & Links to recover removed resources. \\
\bottomrule
\end{tabular}
\caption{Resources used in forum SNCII communities. \textit{Primary resources} are directly used in SNCII creation. \textit{Secondary resources} distribute primary resources and guides for using them. A single submission may contain multiple resources.}
\label{table:resources}
\end{table}

%% file: tables/silver_labels_table.tex
\begin{table*}[t]
\small
\begin{tabular}{llllllll}
\hline
\textbf{Category}         & \textbf{Label}         & \textbf{Count} & \textbf{Percentage} &                       & \textbf{Resource Label} & \textbf{Count} & \textbf{Percentage} \\ \hline
\multirow{11}{*}{Actions} & \textsc{share\_content}         & 111,470        & 39.51               & \multicolumn{1}{l|}{} & \textsc{lora}                    & 13,834         & 4.90                \\
                          & \textsc{discuss}                & 101,958        & 36.14               & \multicolumn{1}{l|}{} & \textsc{model}                   & 10,009         & 3.55                \\
                          & \textsc{react\_praise}          & 21,662         & 7.68                & \multicolumn{1}{l|}{} & \textsc{prompt}                  & 6,209          & 2.20                \\
                          & \textsc{request\_generation}    & 17,702         & 6.27                & \multicolumn{1}{l|}{} & \textsc{ui}                      & 4,088          & 1.45                \\
                          & \textsc{meta}                   & 7,049          & 2.50                & \multicolumn{1}{l|}{} & \textsc{file\_sharing\_service}  & 4,014          & 1.42                \\
                          & \textsc{request\_resource}      & 4,975          & 1.76                & \multicolumn{1}{l|}{} & \textsc{hardware\_instructions}  & 2,304          & 0.82                \\
                          & \textsc{react\_critique}        & 3,754          & 1.33                & \multicolumn{1}{l|}{} & \textsc{dataset}                 & 1,869          & 0.66                \\
                          & \textsc{respond\_request}       & 3,625          & 1.29                & \multicolumn{1}{l|}{} & \textsc{expert\_genner}          & 1,746          & 0.62                \\
                          & \textsc{share\_resource}        & 1,241          & 0.44                & \multicolumn{1}{l|}{} & \textsc{model\_hosting\_website} & 1,730          & 0.61                \\
                          & \textsc{react}                  & 833            & 0.30                & \multicolumn{1}{l|}{} & \textsc{telegram}                & 1,130          & 0.40                \\
                          & \textsc{repost}                 & 408            & 0.14                & \multicolumn{1}{l|}{} & \textsc{credit\_incentive}       & 862            & 0.31                \\ \cline{1-5}
Actors                    & \textsc{socializer\_unknown}    & 220,170        & 78.03               & \multicolumn{1}{l|}{} & \textsc{rentry\_guide}           & 811            & 0.29                \\
                          & \textsc{genner\_unknown}        & 80,825         & 28.65               & \multicolumn{1}{l|}{} & \textsc{session}                 & 547            & 0.19                \\
                          & \textsc{socializer\_veteran}    & 10,903         & 3.87                & \multicolumn{1}{l|}{} & \textsc{nudification\_product}   & 284            & 0.10                \\
                          & \textsc{requester\_celeb}       & 6,098          & 2.16                & \multicolumn{1}{l|}{} & \textsc{web\_archive}            & 79             & 0.03                \\
                          & \textsc{requester\_information} & 3,741          & 1.33                & \multicolumn{1}{l|}{} & \textsc{bitcoin\_link}           & 9              & \textless{}0.01     \\
                          & \textsc{educator}               & 3,659          & 1.30                & \multicolumn{1}{l|}{} &                         &                &                     \\
                          & \textsc{requester\_irl}         & 190            & 0.07                & \multicolumn{1}{l|}{} &                         &                &                     \\
                          & \textsc{genner\_veteran}        & 45             & \textless{}0.01     & \multicolumn{1}{l|}{} &                         &                &                     \\
                          & \textsc{genner\_irl}            & 30             & \textless{}0.01     & \multicolumn{1}{l|}{} &                         &                &                     \\
                          & \textsc{requester\_veteran}     & 2              & \textless{}0.01     & \multicolumn{1}{l|}{} &                         &                &                     \\ \cline{1-5}
Harm Magnifier               & \textsc{minor}                  & 4,153          & 1.47                & \multicolumn{1}{l|}{} &                         &                &                     \\
                          & \textsc{irl\_targeting}         & 1,259          & 0.45                & \multicolumn{1}{l|}{} &                         &                &                     \\ \hline
                          &                        &                &                     &                       &                         &                &                     \\
                          &                        &                &                     &                       &                         &                &                    
\end{tabular}
\caption{\textit{[4chan]} Silver label distribution (N=282,154 posts).}
\label{table:silver_labels}
\end{table*}


%% file: Sections/results.tex
\renewcommand{\dblfloatpagefraction}{0.9}

\section{Results}\label{sec:results}
In this section, we discuss the stakeholders we identify, the resources they share, and how they share them. Figure~\ref{fig:stakeholder} illustrates stakeholders involved in resource dissemination.\footnote{Although targeted individuals can be considered primary stakeholders, we exclude them from our analysis since we primarily characterize the resource dissemination ecosystem, in which targets do not participate.} Primary stakeholders are actively involved in the nudification process; they include forum community members (see Table~\ref{tab:typology}) and nudification tool providers (e.g., commercial app companies, individuals monetizing resource distribution). Secondary stakeholders enable primary stakeholders by providing access to supporting infrastructure, including forums where content is posted, model hosting and file-sharing services, and workflow tools used in generation pipelines.

\begin{sloppypar}
 We organize Section ~\ref{sec:results} around our research questions. Sections \ref{subsec:sharing} and~\ref{subsec:commercial} address \textbf{RQ}\ref{rq:rq1}, resource sharing on 4chan and Reddit respectively. Section~\ref{subsec:influence} addresses \textbf{RQ}\ref{rq:rq2}, the role of influential stakeholders. Section~\ref{subsec:knowledge-transfer} addresses \textbf{RQ}\ref{rq:rq3}, knowledge dissemination by \textsc{educators}. Section~\ref{subsec:promoters} addresses \textbf{RQ}\ref{rq:rq4} by investigating commercial nudification tool sharing by \textsc{promoters}. 
 \end{sloppypar}

\subsection{4chan}
Table~\ref{table:silver_labels} reports raw counts of actor, action, resource, and harm-magnifier labels derived from the 4chan data knowledge graph (Section~\ref{subsec:KG}; 371,096 edges). Labels are not mutually exclusive: a single post may contain multiple actor, action, and resource types, so counts do not sum to N (282,154). \textsc{Socializers} (defined in Section~\ref{subsec:typology}) represent the largest group with 241,653 posts. This is followed by \textsc{genners}, with 80,856 posts, \textsc{requesters} who seek content with 9,780 posts, and \textsc{educators} who provide instructions and guidance with 3,659 posts. \textasciitilde{}42\% of posts (119,418) include images.\footnote{No images were downloaded or stored on research team servers as part of the research process; we determined whether a comment contained media based on HTML metadata and determined content based on text indicators.} 

\subsubsection{Resources Shared in the Ecosystem}
\label{subsec:sharing} 
\begin{figure}[t]
    \centering
    \includegraphics[width=\columnwidth]{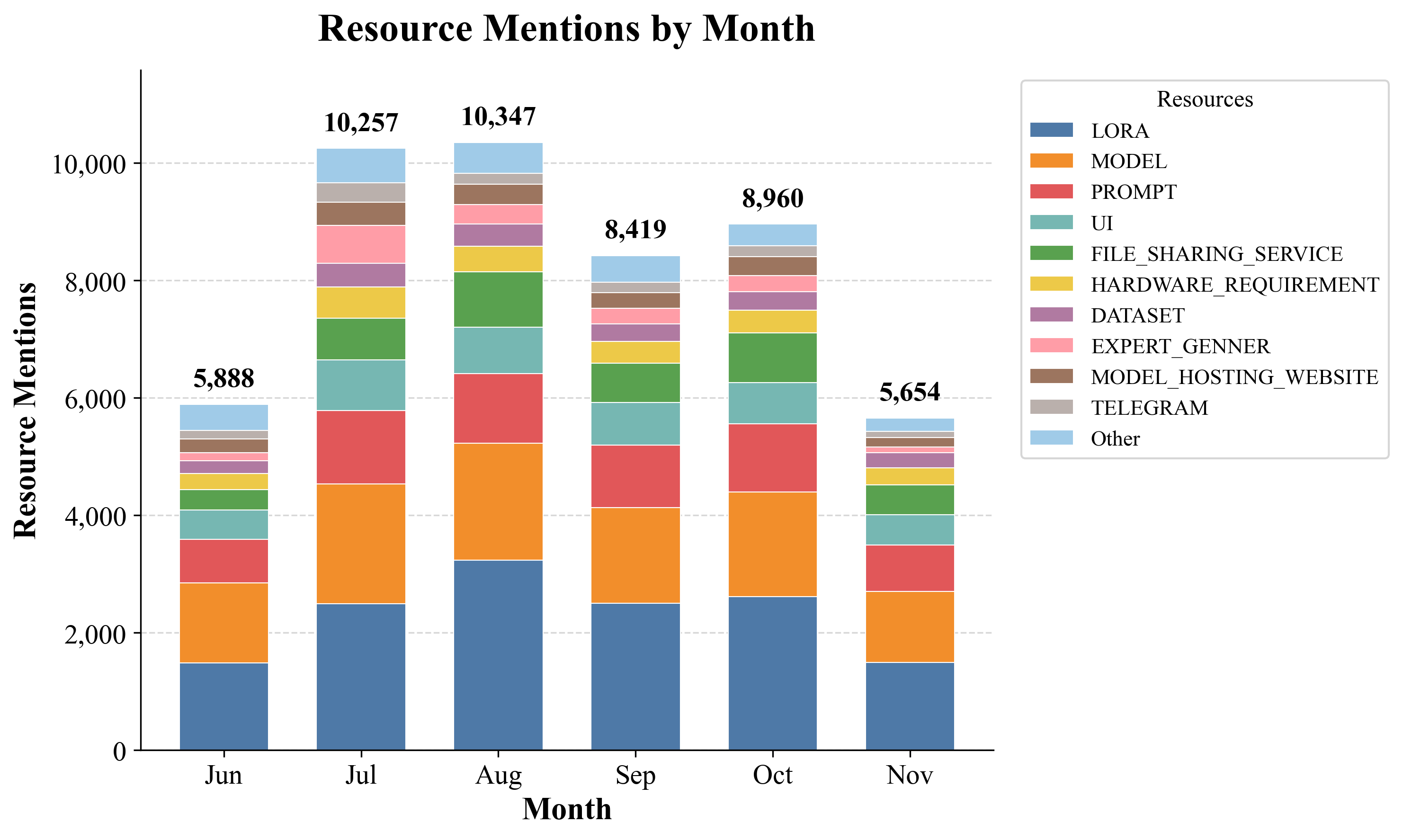}
    \caption{\textit{[4chan]} Resource mentions by month. A brief spike in July-August 2025 follows the removal of real-person likeness models from Civitai in May 2025~\cite{civitai2025likeness}.}
    \label{fig:resource_by_month}
\end{figure}

Tools for independent generation, such as LoRAs, models, prompts, and UI tools, dominate the resource ecosystem, accounting for 75\% of resource requests. Of 282,154 posts, 34,276 (12\%) mention at least one resource. Because individual posts can mention multiple resources, these 34,276 posts collectively yield 49,525 resource mentions across 16 resource types (Table \ref{table:resources}). Of 25,971 request posts, 31\% mention a resource. Demand for resources is concentrated around tools for independent generation: LoRAs, models, and file-sharing services together comprise 67\% of resource requests. Additionally, the general-purpose models Grok and Stable Diffusion account for 12.9\% (1,295) and 11.7\% (1,167) of model mentions, respectively, illustrating that jailbroken general-purpose models represent a significant portion of activity. Notably, resource mentions briefly spike in July-August 2025 following Civitai's removal of real-person likeness models in May 2025 (see Figure~\ref{fig:resource_by_month}). Community members exploit secondary resources to share primary resources, a dynamic illustrated in Conversation \#1, where a user is routed off-platform to find resources:

\noindent\textbf{4chan --- Conversation \#1 --- September 2025}
\begin{list}{}{\leftmargin=1em \rightmargin=1em \itemsep=0.2em \parsep=0pt}
\item \textbf{Comment 1:} Where can I find more LoRAs beyond the link in the original post?
\begin{sloppypar}
\item \textbf{Comment 2:} \emph{(replying to Comment 1)} [expert\_genner] makes some of the best LoRAs but only shares them via torrent because he is American. You can find more on [model\_sharing\_website].
\end{sloppypar}
\end{list}

Our analysis also reveals community opinions about tool efficacy and usefulness. For instance, 89.7\% of reactions to the sharing of a LoRA are positive. Similarly, 94\% of reactions to file-sharing resource posts are positive. File-sharing services are typically used to share LoRAs, datasets of training data, and occasionally SNCII output images. These findings illustrate positive community perceptions of resources facilitating local image generation. In contrast, 30\% of reactions to the sharing of commercial nudification product links are negative. Unlike highly customizable local generation resources such as LoRAs, these commercial nudification products are seen in a more negative light by 4chan users, illustrating the community's predisposition towards certain resource types.

\textit{\textit{Takeaway:}} Model customization tools (LoRAs, models) and distribution channels (file-sharing, Rentry guides) dominate the resource sharing ecosystem  on 4chan, with 75\% of resource requests seeking tools for independent generation. Resource mentions spike only two months after Civitai's removal of real-person likeness models, suggesting possible displacement of SNCII-related resource sharing onto 4chan, with users exploiting secondary resources like model-hosting websites and file-sharing services to disseminate nudification tools among community members~\cite{civitai2025likeness}.

\subsubsection{Influential Stakeholder Roles} 
\label{subsec:influence}
Given the predominantly anonymous nature of the ecosystem (>95\% of posts), we classify influential participants through three lenses: aggregate anonymous activity, individual named user activity, and examination of \textsc{veterans} who engage in meta commentary and show sustained forum presence. We define engagement as the volume of replies, reactions, responses, and discussion posts that a user's content receives. We use engagement and influence interchangeably throughout our analysis.

\textit{Anonymous Influential Activity.}Since anonymous users cannot be tracked individually, we instead characterize influential \textit{activity}. Anonymous users dominate critical ecosystem functions: 94.5\% of \textsc{Genner} posts (76,443 of 80,856), 90.1\% of \textsc{Educator} posts (3,298 of 3,659), and 99.2\% of celebrity request posts (6,052 of 6,098) originate from anonymous users. Harmful content also skews heavily anonymous: 97.1\% (4,031) of minor-related\footnote{Our analysis is only textual; no generated media was viewed. Flagged text mostly references adult public figures in underage scenarios, and targets who recently reached or are soon to reach the age of majority. We reported this thread to the National Center for Missing \& Exploited Children CyberTipline (See~\nameref{sec:ethics}).\label{minornote}} posts and 93.2\% (1,173) of posts targeting private individuals come from anonymous users, suggesting an accountability gap created by platform anonymity. Among the 100 most influential anonymous posts, \textsc{discuss} (51) and \textsc{share\_content} (46) predominate, with LoRAs and model checkpoints being the most frequently shared resources.

\textit{Named Influencers.} To identify behaviors associated with influence, we analyze the posts of the top 10 named users with the highest total engagement (Appendix~\ref{sec:open}). We use the term ``influencers'' to refer to users who receive the highest total engagement across all of their posts. The top 10 influencers contribute 7,929 posts, accounting for 65.7\% of all posts by named users. The most influential named user is a high-volume contributor (2,769 posts), with 56.2\% of their posts consisting of shared generated media (\textsc{Share\_Content}) and 2.2\% of \textsc{Educator} posts. Across the top 10 influencers, every user contributes at least one generation post involving content targeting a minor.

Resource sharing features prominently across top influencers' activity: LoRAs are most frequently shared (797), followed by models (539) and file-sharing service links (307), which typically contain datasets of target media or model checkpoints. Other resources, such as UI tools, prompts, hardware requirements, datasets, model-hosting sites, and Telegram references, appear less frequently. Collectively, these patterns indicate that top contributors facilitate resource circulation, and reuse in the ecosystem.

\begin{figure}[t]
    \centering
    \includegraphics[width=\columnwidth]{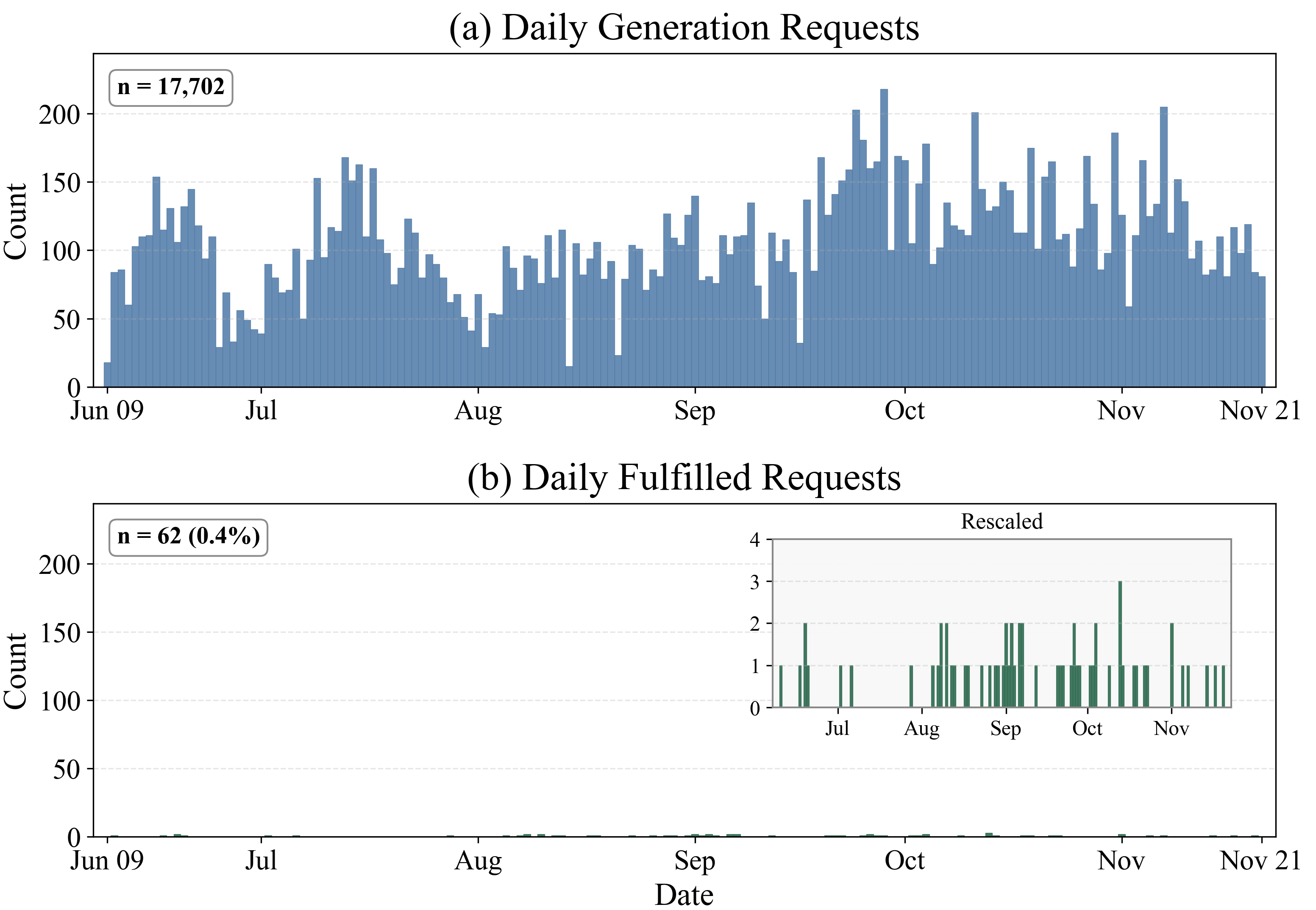}
    \caption{\textit{[4chan]} Monthly generation request volume \& fulfillment (June 9\textendash November 21, 2025). Request rates are high, but fulfillment is very low, suggesting users desire SNCII but may not receive the generations they desire from others.}
    \label{fig:fulfillment}
\end{figure}

\textit{Veterans.} We define \textsc{veterans} as users who either have a username or engage in meta discussions that demonstrate historical ecosystem knowledge. \textsc{Veterans} account for 8\% of total posts (22,568) but represent a larger share of resource-related activity. Specifically, they contribute 13.9\% of posts mentioning file-sharing services (559 of 4,014), 15.3\% of Rentry guide references (124 of 811), 20.0\% of Telegram mentions (226 of 1,130), and 33.8\% of Session~\footnote{Session is a decentralized, end-to-end encrypted messaging application that requires no identifying information to register. Unlike 4chan, where posts are publicly visible and IP addresses are logged by the platform, Session conversations are private and end-to-end encrypted, offering users a significantly higher degree of anonymity.} mentions (185 of 547). 

Manual inspection of Session-related posts indicates that \textsc{veterans} reference off-platform channels to share content considered too extreme for public threads or to reduce perceived legal liability, explicitly referencing U.S. legislation as a motivation. One user notes that they \textit{``share in Session groups since that isn't technically covered in the law,''} while another describes saving \textit{``more extreme''} generations for Session. One of the top 10 named users, discussed in the influencers section above, coordinates private groups, noting that \textit{``I'm in Session now''}. Taken together, these observations suggest that \textsc{veterans} may connect public threads with private channels, where moderation is limited and legal exposure is perceived to be lower. Community members, including \textsc{veterans}, openly observe that 4chan moderation practices further contribute to the ecosystem's persistence, as community members believe they will not be held accountable, encouraging users to continue sharing content and resources despite nominal bans:

\noindent\textbf{4chan --- Conversation \#2 --- November 2025}

\begin{list}{}{\leftmargin=1em \rightmargin=1em \itemsep=0.2em \parsep=0pt}
\item \textbf{Comment 1:} [expert\_genner] should be banned for rulebreaking, but instead, the mods have been banning people who report him.
\item \textbf{Comment 2} \emph{(replying to Comment 1)}: Mods do not do anything here. Session and Kik threads are left up for hours despite being banned.
\end{list}

\begin{figure}[t]
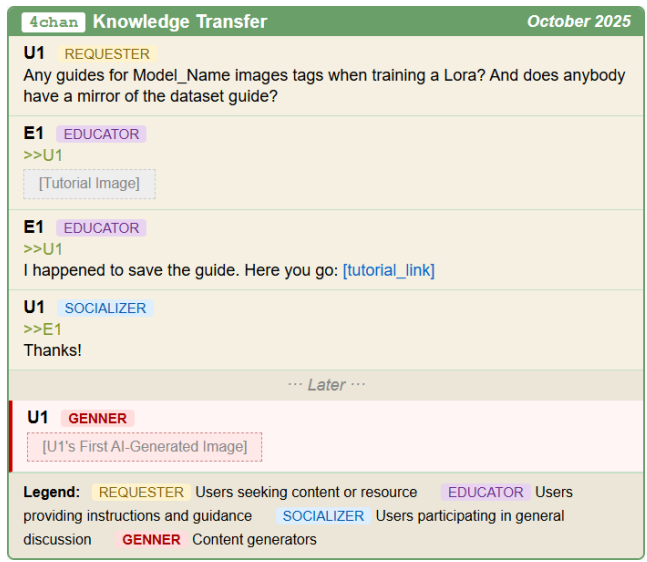

  \small
  \noindent\textbf{4chan --- Knowledge Transfer --- October 2025}

  \begin{description}
    \item[U1 (Requester):] Any guides for \texttt{Model\_Name} image tags
          when training a LoRA? And does anybody have a mirror of the
          dataset guide?
    \item[E1 (Educator), replying to U1:] \emph{[tutorial image]}
    \item[E1 (Educator), replying to U1:] I happened to save the guide.
          Here you go: \emph{[tutorial\_link]}
    \item[U1 (Socializer), replying to E1:] Thanks!
  \end{description}

  \noindent\emph{$\cdots$ Later $\cdots$}

  \begin{description}
    \item[U1 (Genner):] \emph{[U1's first AI-generated image]}
  \end{description}


  \caption{\emph{[4chan]} User U1 first requests guidance on training an
    AI model, prompting three responses from educators offering resources.
    A later post by U1 marks their first instance of SNCII content,
    illustrating how interactions with educators facilitate a shift from
    consumer to creator.}
   \label{fig:case_study}
\end{figure}

\textit{\textit{Takeaway:}} Among named users, influence (engagement) is concentrated. The top 10 influencers by post count hold 61 of the 100 most-engaged posts, with more \textsc{educator} posts (3.66\% vs 0.49\%) and generated content (37.44\% vs 9.91\%) than other named users. \textsc{Veterans} without usernames account for 8\% of posts but play a key role in resource redistribution and off-platform coordination.

\begin{table*}[t]
\centering
\begin{tabular}{lrrrrr}
\toprule
\textbf{Action} & \textbf{Total Posts} & \textbf{Posts Engaged} & \textbf{Total Engagement} & \textbf{Avg} & \textbf{\% Engaged} \\
\midrule
\textsc{Share\_Content} & 111,470 & 63,961 & 105,485 & 1.65 & 57.4\% \\
\textsc{Discuss} & 101,958 & 53,939 & 81,193 & 1.51 & 52.9\% \\
\textsc{Request\_Generation} & 17,702 & 7,365 & 11,576 & 1.57 & 41.6\% \\
\textsc{Meta} & 7,049 & 4,472 & 7,342 & 1.64 & 63.4\% \\
\textsc{Request\_Resource} & 4,975 & 2,734 & 4,300 & 1.57 & 55.0\% \\
\textsc{Respond\_Request} & 3,625 & 2,377 & 4,079 & \textbf{1.72} & \textbf{65.6\%} \\
\textsc{Share\_Resource} & 1,241 & 824 & 1,444 & \textbf{1.75} & \textbf{66.4\%} \\
\bottomrule
\end{tabular}
\caption{\textit{[4chan]} Engagement by Action. \textit{Total Posts}: \# of posts with specific action. \textit{Posts Engaged}: \# of posts receiving at least one interaction. \textit{Total Engagement}: sum of received interactions. \textit{Avg}: average interactions per engaged post (Total Engagement / Posts Engaged). \textit{\% Engaged}: percentage of posts receiving at least one interaction (Posts Engaged / Total Posts $\times$ 100).}
\label{tab:action_engagement}
\end{table*}

\subsubsection{Acquiring SNCII: Requests and Learning.}
\label{subsec:knowledge-transfer}

\begin{sloppypar}
We examine how users seek SNCII through two pathways: requesting content from others and learning to create it themselves.
\end{sloppypar}

\textit{The Request Fulfillment Gap.}  We define a request as fulfilled if it receives a reply containing a shared media or resource. Of 26,523 total requests, only 983 (3.7\%) are fulfilled; 96.3\% remain unanswered. For generation requests (\textsc{Request\_Generation}), the rate is even lower: only 62 of 17,702 requests (0.4\%) receive a response. Figure \ref{fig:fulfillment} shows that fulfillment is very rare: fulfillment rate averages just 0.4\%, with most months having only three fulfilled requests out of 100–200 requests.
This gap illustrates a desire to generate illicit content among users who cannot already do so; this may encourage users either to seek guidance from community \textsc{Educators} in order to become content creators themselves, or to turn to commercial nudification applications that provide the service directly.

\textit{Educator-Driven Role Transitions.}
We observe that interactions with \textsc{educators} are associated with some users subsequently transitioning into content creation roles, even as many requests remain unfulfilled. More than 95\% of the posts are anonymous, limiting role-transition analysis. Our analysis of role transitions therefore focuses on 391 named users who contribute 12,069 posts. Among the 172 named users who post content (\textsc{Genners}), 71 (41.3\%) transitioned from other roles. Of these, 66 (93\%) initially participated in community discussions as \textsc{Socializers}. Notably, 40 of the 71 learned \textsc{Genners} (56.3\%) had interacted with \textsc{Educators} prior to their transition. Figure ~\ref{fig:case_study} presents a representative example. This illustrates that \textsc{Educators} transfer illicit knowledge to \textsc{Socializers} who aspire to generate their own content, facilitating the creation of new \textsc{Genners} in the ecosystem. Users who were already \textsc{Genners} before our data collection began appear as \textsc{Genners} from their first post, so we cannot count them as transitions (i.e., the analysis is left-censored). The 56.3\% figure may therefore be a lower bound.

Knowledge transfer is not limited to the public figure context; some community members express both a desire and ability to nudify private individuals, showing the scope of community targets:

\noindent\textbf{4chan --- Conversation \#3 --- November 2025}

\begin{list}{}{\leftmargin=1em \rightmargin=1em \itemsep=0.2em \parsep=0pt}
\item \textbf{Comment 1:} How can I make realistic NSFW images of people I know in real life?
\item \textbf{Comment 2} (replying to Comment 1): Follow the guide in the original post, but use pictures of the girl you want instead of a celebrity. Following those instructions, I've made LoRAs of every girl I know.
\end{list}

\noindent This example also shows that resources can be shared through ``guides,'' typically hosted on the markdown pastebin service Rentry. Guides contain resources including step-by-step LoRA creation instructions, hyperlinks to model checkpoints and LoRA layers hosted on model-sharing websites or via torrent links, and more. The same guide links are shared each time a new instance of the community thread is created, making them easy to locate. We manually investigate the content of these guides and uncover torrent links to 156 LoRA layers fine-tuned to target 141 female celebrities (with no male targets found), detailed LoRA creation instructions (discussed above), a step-by-step guide to LoRA training using cloud-based GPUs, among other resources. As of submission, many of the guides investigated in our research are no longer accessible, demonstrating the ephemeral nature of the ecosystem.

\textit{Engagement Patterns.} 
We analyze post-to-post interactions across the forum ecosystem to understand what content drives community engagement and to characterize influential users. We define engagement as any observable interaction with a post. Table~\ref{tab:action_engagement} presents engagement metrics by action type. Posts that share content account for the highest total engagement with 105,485 interactions, reflecting their high volume. In contrast, \textsc{Educator} actions such as sharing resources and responding to requests generate higher engagement on a per-post basis. Posts that share resources achieve the highest average engagement at 1.75 interactions per post, and 66.4\% of these posts receive at least one interaction. Responses to requests follow closely, with an average of 1.72 interactions per post and 65.6\% receiving engagement. Overall, while general content sharing dominates in volume, contributions by educators attract high levels of community engagement.

\textit{\textit{Takeaway:}} Generation requests (\textsc{request\_generation}) attract little attention: only 41.6\% receive engagement (lowest among actions) and 0.4\% are fulfilled. In contrast, \textsc{educator} posts receive the highest engagement (1.75 when resource sharing, 1.72 when responding to requests). 40 of 71 new \textsc{genners} (56.3\%) had prior \textsc{educator} interactions, suggesting the ecosystem favors skill transfer over content delivery. 

\subsection{Reddit}
We analyze the 3,683 Reddit submissions that mention a commercial nudification app to characterize app promotion on internet forums.

\subsubsection{Promotion Activity}
\label{subsec:commercial}
We broadly characterize total promotion activity on the six subreddits in our dataset, and promotion over time of the top ten most discussed commercial nudification apps.

\textit{Commercial App Prevalence.} 
Nudification applications are mentioned in all six of the NSFW generative AI subreddits we analyze (see Table~\ref{tab:subreddit-overview}). Smaller communities show higher concentration: up to 41.5\% of submissions in Subreddit \#1 mention these apps. Larger subreddits show lower percentages but higher absolute volumes: Subreddits \#3 and \#6 each contain over 1,000 such submissions. These findings illustrate that \textsc{promoters} use a range of subreddits to attract new users. Notably, ``consent'' is mentioned in only one of the 3,683 total filtered samples, indicating that communicating its importance is not a high \textsc{promoter} priority.

\begin{figure}[t!]
    \centering
    \includegraphics[width=.5\textwidth]{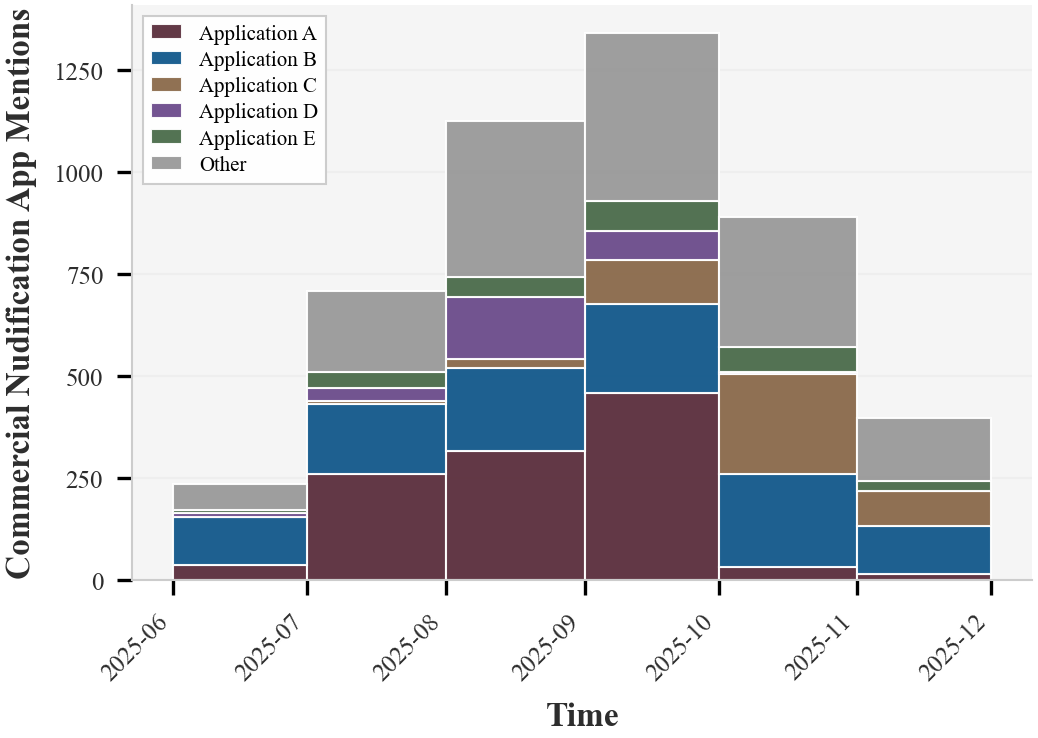}
    \caption{\textit{[Reddit]} Monthly commercial nudification app mentions. Application A mentions dwindle following terms of service changes in October 2025.}
    \label{fig:apps}
\end{figure}

\textit{Longitudinal Analysis by Application.} We analyze application mentions over time to identify longitudinal changes in promotion activity. Figure~\ref{fig:apps} shows promoter mentions of the top five most-mentioned commercial nudification applications over time. Application A dominates mentions until October. Application A's sudden decline in mentions coincides with a change in that application's terms of service (ToS): in October 2025, a clause was added to the ToS indicating that the company reserves the right to terminate any rewards programs and remove any credits from user accounts at their discretion: later in the dataset, most mentions of Application A are of users complaining about having credits removed.

\textit{\textit{Takeaway:}} Up to 41.53\% of submissions contain references to commercial nudification apps, showing that stakeholders are aware of and discuss these apps. Diminishing mentions of Application A corresponding with ToS changes indicate that provider-side action can mitigate dissemination.

\subsubsection{Promoter Activity}
We examine the top 10 most prolific and most influential individual \textsc{Promoters} on our target subreddits to characterize granular promotion activity.

\label{subsec:promoters}
\textit{Referral Links by User.} Of the 17,386 users across the six subreddits, 1,543 (8.9\%) mention at least one commercial nudification app. Among users who mention an app, 1,021 (66.17\%) also share a referral link in at least one of those submissions. Posts with referral links constitute 66.7\% (2,456 submissions) of our full filtered dataset. This indicates that users who discuss commercial nudification apps are also likely to share referral codes.

Across the top 10 most active promoters on our analyzed subreddits, most \textsc{promoters} share referral links across multiple subreddits, and those who share promotions in only one subreddit do so in Subreddit \#6, the largest in our dataset (see Table~\ref{tab:subreddit-overview}). 
This pattern suggests \textsc{promoters} prioritize reach over community focus. Most of the top 10 promoters also share referral links to more than one app, showing that it is common for \textsc{promoters} to use and advertise more than one commercial nudification app. \textsc{Promoter} activity is also short-lived: all top 10 \textsc{promoters} post referral links on fewer than 15 total days (9\% of the 165 days in our dataset); many \textsc{promoters} share links on a single day, and the average promoter posts referral links only 2.4 times.

\begin{table}[t!]
\centering
\begin{tabular}{lrr}
\toprule
\textbf{Subreddit} & \textbf{Submissions} & \textbf{Mentions (\%)} \\
\midrule
\#1 & \textless 1,000 & 41.53 \\
\#2 & \textless 100 & 24.56 \\
\#3 & \textless 15,000 & 12.37 \\
\#4 & \textless 15,000 & 3.97 \\
\#5 & \textless 5,000 & 3.63 \\
\#6 & \textless 50,000 & 2.39 \\
\bottomrule
\end{tabular}
\caption{\textit{[Reddit]} Subreddits Ranked by Percentage of Posts Mentioning Commercial Nudification Apps. \textit{Submissions}: \# of posts and comments on a subreddit, upper-bounded to prevent reidentification. \textit{Mentions}: \# of submissions that mention one or more commercial nudification apps. Percentage is calculated out of the total submissions to that subreddit.}
\label{tab:subreddit-overview}
\end{table}

Often, \textsc{promoters} capitalize on the link-sharing of others:

\noindent\textbf{Reddit --- Post \#1 --- July 2025}

\begin{list}{}{\leftmargin=1em \rightmargin=1em \itemsep=0.2em \parsep=0pt}
\item \textbf{Post:} Here's a link to a Telegram bot for AI undressing. It was used tocreate the attached video.
\item \textbf{Comment 1} (replying to Post): \textit{[referral link]}
\item \textit{[thread has 380 more replies containing referrals]}
\end{list}

\noindent This example illustrates both (a) that Reddit \textsc{promoters} use generated images as advertisements and (b) that there are individual posts with high referral link concentrations, making it easier for new users to find apps. This post receives comments from its initial posting in July all the way until the end of the dataset (November 21), with some repliers replying up to 10 times in quick succession. The high concentration of referral link comments on posts that already contain such a link and the behavioral patterns of repliers suggest that some promotion activity is driven by bot accounts that automatically identify similar activity on target subreddits.

Activity on Reddit is not limited to link-sharing, as users also share issues with commercial products and discuss resources and/or disseminate information to troubleshoot these issues:

\noindent\textbf{Reddit --- Post \#2 --- October 2025}

\begin{list}{}{\leftmargin=1em \rightmargin=1em \itemsep=0.2em \parsep=0pt}
\item \textbf{Post:} Is anyone using \textit{[app1\_name] }on Telegram? I cannot access it anymore. Does anyone know what happened? Is there a backup?
\item \textbf{Comment 1} (replying to Post): I am so upset! I had so many unused credits! Will the app come back? \textit{[app2\_name]} used to come back after being taken down, so hopefully that will happen here as well.
\item \textbf{Comment 2} (replying to Post): I am using this alternative \textit{[referral\_link]}
\end{list}

\noindent The stakeholders in this example are requesting help (\textsc{requesters}) and responding to the request (\textsc{educators}) respectively, suggesting that \textsc{educators} and \textsc{requesters} exist on Reddit, although their activity focuses on commercial apps. Referral links are shared even in troubleshooting discussions, illustrating \textsc{promoter} opportunism.

\textit{Influence.} The bottom-up nature of the promotion process is supported by an analysis of Reddit posters whose submissions receive the most replies: there is no overlap between the top 10 most influential and most prolific posters. The most influential referral link posters are active on average only 3.2 times, and receive an average of 145.3 replies (45.4/post). In contrast, the top 10 most prolific posters post on average 33.8 times (10$\times$ more) but receive only 21 replies (avg. 0.06 replies/submission). Results show that individual \textsc{promoters} are not likely to gain influence through consistent posting. This does not necessarily mean influence on Reddit is inversely correlated with post frequency, but rather indicates that factors beyond post quantity govern influence on Reddit. Of the 2,091 total replies to the top 10 most-replied Reddit submissions containing referral links, 1,026 (49.1\%) of these replies contain hyperlinks to commercial apps. This indicates that \textsc{promoters} take advantage of existing highly influential posts to share their own promotion links. Rather than relying on high individual influence, Reddit promotion relies on multi-level marketing (MLM) dissemination:

\noindent\textbf{Reddit --- Post \#3 --- October 2025}

\begin{list}{}{\leftmargin=1em \rightmargin=1em \itemsep=0.2em \parsep=0pt}
\item \textbf{Post:} \textit{[referral\_link]} share your referral link to earn two free credits when a friend buys credits!
\end{list}

\noindent Similarly, some \textsc{promoters} also share referral links with one another to click each other's referral links and obtain free credits:

\noindent\textbf{Reddit --- Post \#4 --- November 2025}

\begin{list}{}{\leftmargin=1em \rightmargin=1em \itemsep=0.2em \parsep=0pt}
\item \textbf{Post:} Can someone click this link for me? In return, I can click your link. \textit{[referral\_link1]}
\item \textbf{Comment 1} (replying to Post): I clicked it. Click mine \textit{[referral\_link2]}
\item \textit{[all 15 additional replies contain referral links]}
\end{list}

\noindent These examples illustrate not only that \textsc{Promoters} have a financial incentive to recruit new users since they receive credits for doing so, but also that sub-communities of existing \textsc{Promoters} form on Reddit to mutually exploit this MLM-style recruitment strategy; of course, even in attempting to exploit referral links, users end up inadvertently promoting commercial apps by sharing links on public subreddits, showing the effectiveness of this marketing strategy.

\textit{Multi-App Promoters.} A substantial contingent of \textsc{Promoters} (27.2\%, 419) post links to more than one app, with some advertising as many as 11 applications. Similarly, 21.2\% (781) of submissions containing referral links include links to more than one app, with some submissions mentioning as many as 10 apps. These results suggest that promoters either (a) themselves use more than one commercial nudification app, or (b) are paid to advertise more than one service, speaking to the diversity of advertised and used apps. 

We also observe some overlap in referral sharing: 16 codes (1.2\% of codes) spanning multiple apps are posted by 29 different users (2.8\% of referral posters). Low incidence of cross-user referral link reposting might indicate that \textsc{promoter} bans are uncommon and/or that referral links change over time. Indeed, of the top 10 posters, none have been banned from any of the subreddits they post in.

\textit{\textit{Takeaway:}} Referral link sharing is common, with 66.17\% of app mentioners also sharing a referral link. However, Reddit \textsc{promoters} do not reliably accrue influence by posting consistently: because gaining influence is difficult, users sharing referral links have an incentive to share links in reply to already-influential link-sharing posts. These results indicate that credit incentives encourage users who know about/use these apps to share links to them, illustrating the bottom-up nature of promotion activity, and also showing that a broad network of sparsely posting users drive promotion on Reddit.

%% file: Sections/discussion.tex
\section{Discussion}
\label{sec:discussion}
Past work across domains shows that understanding how adversaries operate can inform discussions on how best to combat adversarial ecosystems~\cite{kanich_2008, mirian_2019, franklin2007inquiry, stone_2009}. We apply this framework to the SNCII ecosystem, considering it adversarially from a stakeholder-centric perspective to provide empirical grounding for targeted defense. Focusing on this ecosystem is important because of SNCII's unique negative impacts on psychological safety and individual autonomy~\cite{wei2024understanding, qin2024did}. Below, we synthesize key observations from our findings to support deterrence efforts.

\textit{Ecosystem Overview.} Our empirical analysis uncovers a substantial contingent of stakeholders who frequent the examined forums to discuss and share resources for SNCII creation. We find that a range of relevant primary stakeholders are active in such communities. \textsc{Promoters} who advertise commercial nudification tools account for 4.7\% of submissions to the NSFW AI subreddits we analyze, showing the ease with which one can encounter such resources on the internet if searching for them (Section~\ref{sec:results}). \textsc{Genners}, who actively share illicit content, and \textsc{educators}, who teach others how to create this content, post frequently on 4chan. Unlike on Reddit, the 4chan community prefers to share resources that facilitate local image generation, such as LoRAs, and 4chan reactions to the share of commercial nudification app links indicates that the community frowns on these tools. These findings show that these communities not only exist and perpetrate harm by posting SNCII content and that community members transfer their knowledge to others, growing the contingent of involved users over time even in the wake of regulatory changes (Section~\ref{subsec:knowledge-transfer}), but also that 4chan and Reddit play dual roles in the ecosystem, with different resource types being predominantly shared on each of the two platforms.

\textit{Legal Infrastructure Gap.} The U.S. TAKE IT DOWN Act, passed in May 2025, criminalizes the online distribution of SNCII content. The law also requires ``covered platforms'' (public websites for user-generated content sharing) to create processes allowing victims to request content takedowns~\cite{sen_cruz_s146_2025}. The takedown requirement goes into effect on May 19, 2026, after the end of data collection (although its criminal liability clause is enshrined before this), which may explain the presence of SNCII content on 4chan, but the law's reliance on centralized Federal Trade Commission (FTC) enforcement means there is a risk of selective enforcement~\cite{grimmelmann_deconstructing_2025}. Discussions about the impact of legislation on community practices illustrate that users are cognizant of laws that complicate SNCII content creation. Despite this, users often dismiss these laws' importance:

\noindent\textbf{4chan --- Conversation \#4 --- October 2025}

\begin{list}{}{\leftmargin=1em \rightmargin=1em \itemsep=0.2em \parsep=0pt}
\item \textbf{Comment 1:} What happened with that new US AI law? Did they arrest anyone for their posts?
\item \textbf{Comment 2} (replying to Comment 1): No. Grok and Meta can still make celeb deepfakes. People post them every day on many platforms.
\end{list} 

\noindent This reflects the accountability gap shown in Figure~\ref{fig:stakeholder} from Section~\ref{sec:results}: Users are aware that TAKE IT DOWN only holds liable social media platforms and \textit{sharers} of SNCII, allowing stakeholders like \textsc{educators}, \textsc{promoters}, \textsc{requesters} to continue their activities unhindered. These community responses to accountability gaps demonstrate a willingness to circumvent regulations even as new ones are enshrined, underscoring the community's persistence and accountability; these results align with prior work, which has found that regulations like TAKE IT DOWN only \textit{shifted} users to other platforms~\cite{cuevas2026deepfakepornographyresilientregulatory}. Perceived platform-side apathy (i.e., lack of impetus for 4chan to moderate) further complicates deterrence. 

To our knowledge, no other nation passed new SNCII laws during the data collection period; although the EU has taken steps to ban AI nudifiers under the AI Act, the enshrinement of these measures has been deferred until at least December 2027~\cite{hart_eu_2026}. The lack of recently passed legislation in other jurisdictions may explain TAKE IT DOWN's prevalence in discussions. Yet 4chan users self-identify as posting from the U.S., U.K., and elsewhere, indicating a multinational ecosystem in which users from lax jurisdictions can fill gaps left by stricter ones. For example, the comment below shows a user identifying caveats in their jurisdiction's existing legislation that allow their activity to remain legal:

\noindent\textbf{4chan --- Conversation \#5 --- August 2025}

\begin{list}{}{\leftmargin=1em \rightmargin=1em \itemsep=0.2em \parsep=0pt}
\item \textbf{Comment 1:} The UK Online Safety Act, which I thought was stricter than TAKE IT DOWN, only applies to \textit{[CSAM mention]} and revenge porn. So, technically, celeb AI porn is still legal. They are trying to amend it but UK bros are safe for now.
\end{list}

\noindent These findings align with existing work which has found that heterogeneous regulatory interventions are ineffective deterrents~\cite{cuevas2026deepfakepornographyresilientregulatory}.

\textit{A Decentralized Enabling Infrastructure.} Beyond the forums hosting content, numerous secondary stakeholders contribute to ecosystem persistence (Figure~\ref{fig:stakeholder} from Section~\ref{sec:results}): models are hosted through model-hosting websites and file sharing services (Section~\ref{subsec:sharing}), guides are shared through pastebin services and personal blogs (Section~\ref{subsec:influence}), and general-purpose models and workflow tools are in many cases employed for the creation of SNCII (Section~\ref{subsec:sharing}). When one secondary stakeholder changes its regulations (e.g., when the model hosting service Civitai began removing content representing the likeness of real persons), community members shift to other tools for resource distribution (e.g., file sharing services, see Section~\ref{subsec:sharing}). This adaptive behavior is visible in user discussions:

\noindent\textbf{4chan --- Conversation \#6 --- July 2025}

\begin{list}{}{\leftmargin=1em \rightmargin=1em \itemsep=0.2em \parsep=0pt}
\item \textbf{Comment 1:} I'm downloading a \textit{[female\_celeb\_1]} LoRA. Since Civitai banned celebrity LoRAs, we had to make private new ones, but I can't find LoRAs for \textit{[female\_celeb\_2]} and \textit{[female\_celeb\_3]}.
\end{list}

\noindent This illustrates that fragmented moderation practices contribute to ever-lowering barriers to SNCII creation. Past work on CSAM has spoken to the ineffectiveness of digital self-regulation by platforms absent structural changes addressing the inherent compliance differences introduced by digital content sharing (i.e., territorial jurisdiction complicates enforcement due to the non-territorial nature of digital content sharing)~\cite{bleakley2024moderating}. Aligning with these past findings, our analysis emphasizes the importance of creating a legal infrastructure that not only holds all involved stakeholders appropriately accountable but is also standardized across jurisdictions. 

\textit{Platform Interventions Can Work.} Reddit data shows application-side action can disrupt activity (Section~\ref{subsec:sharing}): Application A's ToS change removing credit incentives led to sharp declines in promotional mentions (Figure~\ref{fig:apps}), though some users migrated to alternatives. On 4chan, sporadic moderation allows content to persist. This contrast suggests platform action is necessary but insufficient without coordination across providers. Recent reports have emphasized the importance of establishing effective deterrence mechanisms, including by platform-level monitoring to raise awareness as to tools used for SNCII distribution~\cite{thakur_think_2025, kamachee_video_2026}. Our work supplements existing recommendations by identifying Reddit as a source for SNCII-enabling app URLs; as such, given our findings in Section~\ref{sec:results}, we propose that regular crawls of Reddit may be effective in identifying content creation and distribution channels.

\textit{Intervention Points.} Building on the chokepoints identified in Gibson et al.'s foundational work on commercial nudification tools~\cite{gibson2025analyzing}, we suggest additional chokepoints identified through our resource- and stakeholder-focused lens. First, \textit{educator disruption}: on 4chan, educators comprise only 1.3\% of posts, yet 56.3\% of users who later became \textsc{Genners} had prior interactions with \textsc{Educators} (Section~\ref{subsec:influence}); this indicates that regulating SNCII \textit{resources} solicitation and distribution may be more effective deterrent than targeting content alone. Second, \textit{infrastructure chokepoints}: payment processors, model hosting platforms, and file-sharing services are critical to the resource distribution process and therefore should be subject to more stringent regulation. Third, \textit{moderation harmonization}: inconsistent enforcement across platforms allows users to move activity to undermoderated spaces (see Section~\ref{subsec:influence}). Fourth, \textit{extended liability}: current U.S. law exempts \textsc{Requesters}, \textsc{Educators}, and tool developers who are essential to ecosystem function, as seen in Figure~\ref{fig:stakeholder} (see Section~\ref{sec:results}). Fifth, \textit{unified liability}: users from jurisdictions with lax SNCII laws can fill gaps left by patchwork enforcement unless policy is harmonized internationally. Our results (spanning Section~\ref{sec:results}) indicate single-point interventions are likely to be insufficient, suggesting coordinated action is essential to reduce SNCII creation activity~\cite{cuevas2026deepfakepornographyresilientregulatory}. Of course, in extending liability, regulators must take care not to create overly restrictive frameworks that generate unintentional harm (take, for instance, online age verification technology, which, while intended to protect children, has had much broader repercussions~\cite{colliver2024porn, stardust2024mandatory}). Past research investigating SNCII has found that deplatforming alone merely causes content to shift to other platforms, emphasizing the importance of a harmonized regulatory structure~\cite{cuevas2026deepfakepornographyresilientregulatory}.

While adversarial practices surrounding SNCII creation will continue to evolve, the underlying behaviors we document (i.e., sharing of both easy-to-use and high-sophistication tools, and adapting content creation activities to conform to the letter of the law) are likely to persist. Thus, while our work is grounded in specific datasets, we believe that the patterns we identify are likely to persist as technologies evolve. As these communities continue to move further underground, a practice that is already happening and speaks to the ecosystem's high level of adaptability, such analyses may become both inaccessible to empirical inquiry and increasingly relevant for understanding how these ecosystems sustain themselves.

\textit{Dataset Limitations \& Future Work.} We focus our analysis on Reddit and 4chan because these forums are well-known and have been reported as relevant to the SNCII creation ecosystem~\cite{ding_malicious_2025}. Within 4chan, we only analyze the /b/ board. SNCII activity also occurs on other boards, including the adult /r/ board, but the archive websites hosting threads from that board explicitly prohibit the use of automated tools for data gathering. To avoid violating terms of service, we constrain our analysis to /b/. The resources and actors we characterize are therefore a subset of SNCII activity on 4chan, though our findings also show that this activity is not limited to the adult /r/ board. Actors/actions/resources supplementing those we describe may be shared on other internet forums or through other distribution channels. These channels should be explored by future work. Further characterization of resources uncovered by this work (e.g., Session, Rentry guides) is also an area to be explored by future research. For Reddit, other subreddits beyond the six NSFW AI subreddits we examine may contain discussion of additional commercial apps or other resources.

%% file: Sections/related_work.tex
\section{Related Work}
\label{sec:rel-work}
Prior work has focused on how individuals experience and interpret SNCII harms and the accessibility of generating these images. Studies of user perceptions emphasize the privacy violations caused by SNCII, documenting the ways victims conceptualize harm and the social consequences of abuse~\cite{brigham2024violation,umbach2024non}. In parallel, research has also examined how affected individuals navigate online spaces for support and information~\cite{wei2024understanding}. This work highlights how help-seeking behaviors emerge on the internet in response to SNCII, underscoring that the same platforms that enable harm may also serve as critical venues for recovery and assistance. Some research has examined the motivations of perpetrators through qualitative strategies, revealing a broad range of motivations for stakeholder participation in SNCII ecosystems~\cite{mink2026unlimited}. Other research has considered teachers' perspectives on SNCII generation by students~\cite{wei2025students}.
Prior work also shows that synthetic images can be generated even with limited computer skills~\cite{Mehta_2023}, lowering the barrier to producing SNCII content. Collectively, this work provides important context on the human aspects and impacts of SNCII and motivates the need for interventions that address both individual and systemic factors.

More broadly, adversarial behavior has been studied across a range of online abuse contexts, demonstrating that such communities rely on shared tooling, infrastructure, and coordination. Measurement studies of underground economies show how abuse becomes commoditized through markets for spamming, phishing, credential theft, and compromised hosts~\cite{kanich_2008, mirian_2019, franklin2007inquiry, stone_2009}. Analyses of underground forums further reveal social structures in which trust is built and illicit goods and services are exchanged, illustrating how abusive ecosystems organize and sustain themselves~\cite{motoyama2010re}. Other work shows how defensive mechanisms can give rise to service ecosystems, such as CAPTCHA-solving markets that combine automated tools and human labor to bypass platform protections and frame abuse in economic terms~\cite{motoyama_2011}. Complementary studies quantify the economic viability of large-scale abuse and show how affiliate programs and service-oriented markets enable coordination across roles and shared infrastructure, even in the presence of low conversion rates~\cite{kanich_2008, mccoy_2012, mirian_2019}. Together, these works motivate ecosystem-level analysis of adversarial communities, but do not capture the distinct dynamics introduced by SNCII.

%% file: Sections/conclusion.tex
\section{Conclusion}
\label{sec:concl}
This work presents the first stakeholder-centric characterization of SNCII community practices on internet forums. As discussed in Section~\ref{sec:discussion}, studying these ecosystems provides an informed understanding of current practices and, hence, a foundation upon which to consider approaches for combating SNCII creation and dissemination. As such, we perform an analysis of 282,154 4chan posts and 78,308 Reddit submissions, revealing a complex ecosystem sustained by distinct stakeholder roles operating across platforms with varying levels of technical sophistication.
Our analysis identifies intervention points that include educator disruption, infrastructure choke points, and extended and unified liability frameworks. Further, the predominantly ephemeral nature of the ecosystem creates fundamental accountability challenges, underscoring the importance of harmonized enforcement. Effective intervention will require coordinated action targeting not only content sharers, but the broader ecosystem enabling SNCII creation.

%% file: Sections/ethics.tex
\phantomsection
\section*{Ethical Considerations}
\label{sec:ethics}
\textit{Forum Research Ethics.} Researching online forum communities comes with unique ethical implications~\cite{fiesler_remember_2024}. We take steps to ensure that our research process is consistent with the ethical considerations recommended by~\cite{fiesler_remember_2024}: we are mindful of the examined communities (especially on Reddit, where harmful content is a subset of the full dataset), and take steps to avoid drawing undue attention to examined communities by not naming them. The presence of illicit content on the forums being studied motivates our investigation, as understanding harmful ecosystems such as these is key to identifying chokepoints for moderation. Although the forums being investigated are public, we acknowledge that users may not wish for their posts to be quoted in an academic work, and as such, we take steps to ensure no data is identifiable. In particular, all quoted posts are paraphrased, and an author verified that the paraphrased posts were not reverse-searchable. Furthermore, no usernames are reported. These measures are in line with established best practices for research with public internet data~\cite{kohno2023ethical, bruckman2002studying}.

\textit{Scraping.} We acknowledge that scraping SNCII forum data involves some interaction with harmful communities. To mitigate risks, we scrape an archive website for data (in the case of 4chan) and use dumped data accessible via an API created by researchers (in the case of Reddit). As such, we never directly interact with or contribute to the traffic of any illicit forum communities. We constrain our scraping to 4chan's /b/ community as archive websites hosting content from other communities explicitly prohibit scraping in their terms of service. As such, no terms of service were violated in the data collection process. Additionally, no images are saved in the process of obtaining our data.

\textit{Data Anonymization.} We recognize that work in this space has the potential to guide malicious actors to resources that might allow them to engage in illicit and/or illegal behavior. To mitigate this risk, we anonymize references to named users, applications, and models throughout our work. Multiple researchers reviewed the work to ensure that no identifiable information that might impose real-world harm (i.e., by teaching an adversary how to generate such images themselves) was included in the work. Data anonymization practices are also in place to ensure that no individual targets of SNCII content found in our analysis are identifiable in this work, so as not to draw attention to any individual and minimize the risk of imposing psychological harm. Similarly, all user quotations included in the work are paraphrased to prevent reidentification of individual users or threads.

\textit{Prominent Stakeholders.} We thought carefully about whether to name prominent stakeholders in our paper. We ultimately decided it was appropriate to name stakeholders to establish context for this work and contextualize resource-sharing; since the stakeholders we name are largely well-known and many have been covered by mainstream media, we ultimately decided that naming these stakeholders was not likely to cause harm.

\textit{Reporting.} Posts containing text that references or targets minors were identified during our analysis. We reported the 4chan thread where these posts were found to the CyberTipline of the National Center for Missing \& Exploited Children
(\url{https://www.missingkids.org/}) to support content identification and removal.

\textit{Researcher Well-Being.} This work required direct engagement with disturbing text content by researchers. The researchers ultimately decided that this exposure was worthwhile and necessary to expand our understanding of the resources employed by these illicit communities. To minimize researcher exposure to traumatic content, we opted to exclude image-level analysis, instead analyzing only sample text and a boolean variable indicating whether or not an image was included. Researchers also took breaks as needed throughout the data analysis process.

%% file: Sections/open-science.tex
\makeatletter
\section*{Appendix}
\section{Open Science}
\label{sec:open}
We acknowledge the importance of reproducibility in the scientific community. Therefore, we provide additional methods detail and our full codebook \& dataset labeling statistics as supplementary materials. We also provide: (a) a detailed codebook (\texttt{codebook\_detailed.html}) with granular actor, action, and resource categories and code selection instructions; (b) an annotation tool and example json file, so researchers may apply codebook to similar data; (c) scripts for merging annotations with source data and importing the dataset into a Neo4j knowledge graph; and (d) a list of the queries used to extract quantitative insights, along with a script to execute them, so that researchers may use analogous queries to extract comparable information. These artifacts support the paper's core contributions: the multi-label behavioral codebook, the annotation pipeline, and the knowledge-graph-based analytical framework. All materials are available at: \url{https://doi.org/10.5281/zenodo.22694536}. 

To minimize the risk of creating harm by allowing malicious actors to track down illicit tools through open-sourced data, we opt not to share our full datasets or additional info about the examined communities. Releasing this data would effectively function as a directory of illicit resources and could compromise the safety of targeted individuals. To this end, we share our codebook, pipeline, and queries to support the evaluation of our methodology and to enable researchers to apply our approach to comparable data. However, we will share upon request full datasets with researchers at academic institutions seeking to conduct future work. In cases such as this one, where research involves extremely sensitive data, vetting minimizes the creation of harm while maximizing the potential for data sharing between members of the academic community.

%% file: Sections/appendix.tex
\section{Methodology}
\label{sec:methodology-appendix}

We provide here further detail on our data collection and analysis methodologies for reproducibility purposes.

\subsection{Data Collection}
\label{sec:data-collection-appendix}

\paragraph{4chan} The 4chan community we analyze is limited to one thread: the group creates a single thread for image and resource sharing, and once it hits 4chan's post limit, they create a new one. Thread creators often link to the previous thread, which we used to chain our collection. Collection begins June 9th (earliest archived data) and ends November 21st when the community migrated to another board (/r/ the ``adult'' community) on this day, and this board is not archived on the same website from which we scrape our data.

\subsection{Data Analysis}
\label{sec:data-analysis-appendix}
\subsubsection{Preliminary Characterization}
\label{subsubsec:char}
Through our preliminary exploration, we discovered that the resource sharing behaviors of prevalent stakeholders differ between the two communities we study: 4chan stakeholders prominently discuss publicly downloadable model variants~\cite{hawkins2025deepfakes} and jailbroken general-purpose models, whereas resources shared on Reddit are frequently commercial tools~\cite{gibson2025analyzing}. Furthermore, while 4chan activity is exclusively targeted towards SNCII since the thread we study is created specifically for this purpose, Reddit submissions are noisier, with some posters discussing AI-generated sexual content of fictional characters (which is out of scope).  Differing resource sharing practices reflect the prevalence of different actors in each forum: \textsc{Genners} and \textsc{Educators} dominate on 4chan, while most resource sharing activity observed on Reddit is conducted by \textsc{Promoters}. These findings illustrate that the datasets we examine are complementary. 

We recognize that Reddit communities might contain topical discussions beyond the promotion of commercial nudification apps. However, we found through our preliminary exploration that much of the discussion on our target subreddits not related to commercial nudification apps was about AI generation of sexual content not targeting real people; since this content is not in the scope of the threat model we consider, and since filtering of harmful vs. not harmful sexual content on Reddit would require complex filtering that is beyond the scope of this work, we opt to constrain our analysis of Reddit to commercial nudification applications. 

\subsubsection{4chan Automated Analysis}
\label{subsubsec:4chan-automated}
\paragraph{SetFit Classification} For actors and actions, we train SetFit~\cite{tunstall2022efficient}, a sentence-transformer-based few-shot classifier, using a one-vs-rest multi-label strategy, training for 10 epochs with batch size 32. We choose SetFit because these labels depend on communicative intent, not surface-level keywords.

\paragraph{N-grams} For resources and harm magnifiers, we extracted discriminative n-grams from the 400 annotated samples. For each category, we scored candidate n-grams using a TF-IDF-weighted uniqueness metric: $\text{score} = \text{tf} \times \text{idf} \times \text{uniqueness}$, where uniqueness measures what fraction of an n-gram's occurrences appear in that category. We retained the top 20 patterns per category with minimum count $\geq 2$ and uniqueness $\geq 0.3$. Patterns include tool names (e.g., ``ComfyUI''), platform domains, and community jargon. These were converted to case-insensitive regex rules with word boundary matching. Performance differences (F1=0.68 for actions vs F1=0.92 for resources) reflect the distinction between categories requiring intent interpretation and those with clear surface signals.

\paragraph{Knowledge Graph Construction} We construct $G(V, E)$ where $V$ represents posts and $E$ captures reply-based interactions. Each comment becomes a \textbf{Post} node carrying metadata and labels as attributes. Since multiple actions can occur in a single post, nodes can have multiple action and actor labels. When a post references another using 4chan's reply markup (\texttt{>>post\_id}), we create edges based on the replying post's properties. Two nodes may share multiple edges capturing distinct interaction types.

\subsubsection{Reddit Manual Analysis} 
\label{subsubsec:reddit-manual}
\paragraph{Building Keyword Filter} One researcher reviewed 400 Reddit submissions, identifying 256 samples that contain commercial nudification app mentions. The 53 unique application names extracted from these samples comprise our final keyword filter. A second researcher reviewed the curated samples and the final keyword filter list to ensure accuracy.

\subsubsection{Reddit Automated Analysis} 
\label{subsubsec:reddit-automated}
\paragraph{URLs} To characterize commercial application prevalence, we measure the proportion of posts containing URLs to commercial nudification apps by searching for posts that contain URLs and using our keyword filter list to remove URLs that do not refer to a commercial nudification app. We also examine the proportion of unique referral codes shared in hyperlinks by searching filtered hyperlinks for common referral code patterns (e.g., ``r='', ``ref=''). 

\paragraph{Users \& Apps} We supplement our broader analysis with user-level and application-level analysis. We identify the most active posters of referral links by mapping posted links to the users who posted them, the date of the post, the number of subreddits the user is active in, and the number of days they are active for. We also examine mentions of individual commercial nudification applications over time by identifying the posts in our full filtered dataset that contain one or more mentions of at least one target app.

\clearpage

\section{Tables}
\label{sec:tables}
Granular information about individual posters is given in Tables~\ref{tab:top10_profiles} and~\ref{tab:top-referral-users}. We report our full codebook in Table~\ref{table:codebook}. 

\input{tables/4chan_influencers.tex}
\input{tables/reddit_top-poster_table.tex}
\input{tables/codebook_table.tex}

%% file: tables/4chan_influencers.tex
\begin{table}[h!]
\centering
\resizebox{\columnwidth}{!}{
\begin{tabular}{@{}rrrrrrr@{}}
\toprule
\textbf{Rank} & \textbf{Posts} & \textbf{Engagement} & \textsc{\textbf{Educator}} & \textsc{\textbf{Genner}} & \textsc{\textbf{Minor}} \\
\midrule
1 & 2,769 & 9,419 & 61 & 1,556 & 15 \\
2 & 1,011 & 2,956 & 41 & 222 & 3 \\
3 & 944 & 925 & 87 & 163 & 11 \\
4 & 859 & 1,376 & 4 & 268 & 2 \\
5 & 537 & 2,287 & 17 & 149 & 4 \\
6 & 506 & 1,170 & 6 & 253 & 3 \\
7 & 459 & 560 & 1 & 35 & 11 \\
8 & 358 & 1,035 & 64 & 162 & 3 \\
9 & 276 & 824 & 8 & 130 & 1 \\
10 & 210 & 779 & 1 & 31 & 5 \\
\bottomrule
\end{tabular}
}
\caption{\textit{[4chan]} Top 10 4chan named influencer profiles.}
\label{tab:top10_profiles}
\end{table}

%% file: tables/reddit_top-poster_table.tex
\begin{table}[h!]
\centering 
\begin{tabular}{cccccc}
\toprule
\textbf{Rank} & \textbf{Activity} & \textbf{Ref (\%)} & \textbf{Subreddits} & \textbf{Apps} & \textbf{Days} \\
\midrule
1 & 88 & 100 & \#6 & a, b & 8 \\
2 & 42 & 100 & \#6 & a, b & 1 \\
3 & 34 & 69.4 & \#3, \#4, \#6 & a, b & 14 \\
4 & 31 & 75.6 & \#4, \#5, \#6 & a, b & 8 \\
5 & 26 & 100 & \#3, \#4, \#6 & a, b & 10 \\
6 & 26 & 96.3 & \#3, \#6 & a & 5 \\
7 & 24 & 100 & \#6 & a & 4 \\
8 & 23 & 76.7 & \#3, \#4, \#6 & c, d & 2 \\
9 & 22 & 100 & \#3, \#6 & a, b, e & 3 \\
10 & 22 & 100 & \#6 & a & 1 \\
\bottomrule
\end{tabular}
\caption{\textit{[Reddit]} Top \textsc{promoters} by Referral Link Posting Activity.  \textit{Activity}: total \# of submissions containing a referral link. \textit{Ref (\%)}: total percent of activity made up by referral link sharing. \textit{Days}: \# of days on which the user posted.}
\label{tab:top-referral-users}
\end{table}

%% file: tables/codebook_table.tex
\begin{table*}[h]
\centering
\begin{tabular}{@{}llr|llr@{}}
\hline
\textbf{Code Category} & \textbf{Code} & \textbf{Count} & \textbf{Code Category} & \textbf{Code} & \textbf{Count} \\
\hline
\textbf{\textit{Actor}} & \textbf{Socializer} & 290 & \textbf{\textit{Resource}} & \textbf{Resource Type} & \\
& \quad socializer & 223 & & \quad lora & 69 \\
& \quad socializer\_veteran & 67 & & \quad model & 49 \\
& \textbf{Genner} & 134 & & \quad file\_sharing\_service & 28 \\
& \quad genner\_unknown & 64 & & \quad ui & 19 \\
& \quad genner\_unknown\_veteran & 35 & & \quad prompt & 19 \\
& \quad genner\_celeb & 19 & & \quad model\_hosting\_website & 12 \\
& \quad genner\_celeb\_veteran & 12 & & \quad hardware\_requirement & 12 \\
& \quad genner\_irl & 4 & & \quad expert\_genner & 11 \\
& \quad genner\_irl\_veteran & 0 & & \quad dataset & 11 \\
& \textbf{Requester} & 112 & & \quad telegram & 10 \\
& \quad requester\_celeb & 39 & & \quad nudification\_product & 7 \\
& \quad requester\_irl & 1 & & \quad session & 5 \\
& \quad requester\_celeb\_veteran & 3 & & \quad rentry\_guide & 5 \\
& \quad requester\_irl\_veteran & 0 & & \quad credit\_incentive & 4 \\
& \quad requester\_resource & 56 & & \quad bitcoin\_link & 2 \\
& \quad requester\_resource\_veteran & 5 & & \quad web\_archive & 2 \\
& \quad requester\_target\_unknown & 12 & & \textbf{Sharing Type} & \\
& \textbf{Educator} & 47 & & \quad mention & 132 \\
& \quad educator & 47 & & \quad link & 31 \\
& & & & \quad instructions & 23 \\
\textbf{\textit{Action}} & \textbf{Discuss} & 182 & & \quad review & 9 \\
& \quad discuss & 182 & & & \\
& \textbf{Meta} & 50 & \textbf{\textit{Harm Magnifier}} & \quad minor & 19 \\
& \quad meta & 50 & & \quad irl\_targeting & 10 \\
& \textbf{React} & 120 & & & \\
& \quad react & 44 & & & \\
& \quad react\_praise & 52 & & & \\
& \quad react\_critique & 25 & & & \\
& \textbf{Respond} & 48 & & & \\
& \quad respond\_request & 48 & & & \\
& \textbf{Repost} & 3 & & & \\
& \quad repost & 3 & & & \\
& \textbf{Advertise} & 6 & & & \\
& \quad advertise\_recommend & 6 & & & \\
& \textbf{Share} & 147 & & & \\
& \quad share\_content & 123 & & & \\
& \quad share\_resource & 32 & & & \\
& \quad share\_experience & 10 & & & \\
& \textbf{Request} & 113 & & & \\
& \quad request\_resource & 61 & & & \\
& \quad request\_generation & 54 & & & \\
\hline
\end{tabular}
\caption{Actor, Action, Resource, and Harm Magnifier Codes (n=400 4chan comments).}
\label{table:codebook}
\end{table*}